\documentclass[twocolumn, twocolappendix]{aastex63}

\usepackage{times}

\usepackage{graphicx}

\usepackage{tikz}

\usepackage[T1]{fontenc}
\usepackage{aecompl}
\usepackage{calligra}
\DeclareMathAlphabet{\mathcalligra}{T1}{calligra}{m}{n}
\DeclareFontShape{T1}{calligra}{m}{n}{<->s*[2.2]callig15}{}

\usepackage{hyperref}
\hypersetup{
    pdfnewwindow=true,      
    colorlinks=true,       
    linkcolor=orange,      
    citecolor=cyan,        
    filecolor=orange,      
    urlcolor=orange           
}

\usepackage{color}

\newcommand{\scripty}[1]{\ensuremath{\mathcalligra{#1}}}

\def\scr{\scripty{r}}

\def\max{\mathrm{max}}

\def\Mach{\mathcal{M}}
\def\Hside{\Theta}

\def\adot{\dot{a}}
\def\edot{\dot{e}}
\def\Mdot{\dot{M}}

\def\epsdot{\dot{\epsilon}}
\def\ldot{\dot{l}}

\def\disco{\texttt{DISCO}~}

\def\eg{\textit{e.g.}}

\begin{document}

\shorttitle{Eccentric Binary Orbital Evolution}
\shortauthors{D'Orazio $\&$ Duffell}

\title{Orbital Evolution of Equal-mass Eccentric Binaries due to a Gas Disk: \\Eccentric Inspirals and Circular Outspirals}

\author[0000-0002-1271-6247]{Daniel J. D'Orazio}
\affiliation{Niels Bohr International Academy, Niels Bohr Institute, Blegdamsvej 17, 2100 Copenhagen, Denmark}
\email{daniel.dorazio@nbi.ku.dk}

\author[0000-0001-7626-9629]{Paul C. Duffell}
\affiliation{Department of Physics and Astronomy, Purdue University, 525 Northwestern Avenue, West Lafayette, IN 47907-2036, USA}

\begin{abstract}
We solve the equations of two-dimensional hydrodynamics describing a circumbinary disk accreting onto an eccentric, equal-mass binary. 
We compute the time rate of change of the binary semimajor axis $a$ and eccentricity $e$ over a continuous range of eccentricities spanning $e=0$ to $e=0.9$. We find that binaries with initial eccentricities $e_0\lesssim0.1$ tend to $e=0$, where the binary semimajor axis expands. All others are attracted to $e\approx0.4$, where the binary semimajor axis decays. The $e\approx0.4$ attractor is caused by a rapid change in the disk response from a nearly origin-symmetric state to a precessing asymmetric state. The state change causes the time rates of change $\adot$ and $\edot$ to steeply change sign at the same critical eccentricity resulting in an attracting solution where $\adot = \edot = 0$. This does not, however, result in a stalled, eccentric binary. The finite-transition time between disk states causes the binary eccentricity to evolve beyond the attracting eccentricity in both directions resulting in oscillating orbital parameters and a drift of the semimajor axis. For the chosen disk parameters, binaries with $e_0\gtrsim0.1$ evolve toward and then oscillate around $e \approx 0.4$ where they shrink in semimajor axis. Because unequal mass binaries grow toward equal mass through preferential accretion, our results are applicable to a wide range of initial binary mass ratios. Hence, these findings merit further investigations of this disk transition; understanding its dependence on disk parameters is vital for determining the fate of binaries undergoing orbital evolution with a circumbinary disk. 
\end{abstract}

\keywords{hydrodynamics --- binaries: general --- stars: formation --- accretion, accretion disks --- quasars: general --- galaxies: active -- gravitational waves}

\section{Introduction}
\label{s:Introduction}
The interaction of a binary and a gas disk arises in a wide range of astrophysical scenarios. Namely, the birth and evolution of stellar and planetary systems \citep[\eg,][]{Mathieu_DQTau+1997, Alves+2019, DMartin_CBplanets:2019}; formation channels for compact-object-binary mergers \citep{Tagawa+2020, LiDempsey+2021} presently being detected in gravitational waves \citep{LIGO_GWTC1:2019}; and the hardening of supermassive black hole binaries and the final parsec problem \citep{Begel:Blan:Rees:1980, GouldRix:2000, ArmNat:2002}, crucial for understanding the low frequency gravitational-wave sky including the gravitational-wave background probed by the Pulsar Timing Arrays \citep{NANOGrav12p5_2020}, and the supermassive black hole binary merger rate by LISA \citep{LISA:2017}.

The gas-disk interaction dictates not only electromagnetic signatures of the binary \citep[\eg,][]{HKM09, PG1302MNRAS:2015a, Tofflemire_TWA3A+2017}, but also its orbital evolution. The former provides a means for observational identification and system characterization, while the later is a main ingredient in population synthesis schemes needed to understand observed populations \citep[\eg,][]{El-BadryTwins+2019, AMPW_APOGEEII+2020} as well as extrapolate to the undiscovered \citep[\eg,][]{Kelley+2019}.

Here we focus on binary orbital evolution in the limit of a thin circumbinary disk accreting onto an equal-mass, eccentric binary. The majority of recent work in this effort has focused on circular orbit binaries, measuring relative accretion rates onto binary components and the gas-induced torque on the binary as a function of mass ratio. Recently, a commonly held picture of gas driving binaries toward inspiraling circular orbits \citep{ArmNat:2002, MM08} has been called into question. This is due to a number of works carrying out high resolution simulations and enacting a careful analysis of angular momentum transport through the disk onto the binary to find {\em expanding} binary orbits
\citep{MirandaLai+2017, Tang+2017, MunozLai+2019, MoodyStone:2019, Duffell+2020, Munoz_FinDsk+2020}, although dependence of these results on hydrodynamic parameters is less explored \citep[but see][]{Tiede+2020, HeathNixon:2020, Duffell+2020, Munoz_FinDsk+2020}. 
In particular, \citet{Duffell+2020} (hereafter DD20) study the mass accretion and torque on circular orbit binaries with mass ratios continuously spanning the range of $1:100$ to unity. DD20 shows that preferential accretion (as found in a number of works, \eg,
\citet{Bate+2000, Farris+2014}) acts to drive all such binaries toward $1:1$ mass ratios, but that only binaries below mass ratios of $\sim1:20$ are driven together; above this, binaries expand outwards.\citet{Munoz_FinDsk+2020} present a similar picture finding expanding orbits for $q\gtrsim0.2$.

How does orbital eccentricity evolution affect this picture? While previous work has suggested that binary eccentricity growth may be important \citep[][]{Roedig+2011, MirandaLai+2017, MunozLai+2019}, only recently did \citet{Zrake+2020} measure the time rate of change of eccentricity for equal-mass binaries over a large range of eccentricities, finding that initially low eccentricity systems tend toward circular orbits, while higher initial-eccentricity systems tend toward eccentricities of $e\sim0.45$. That work did not explore binary semimajor-axis evolution, however, and hence, leaves the fate of such binaries uncertain.

In this {\em Letter} we carry out high resolution hydrodynamical calculations to measure gas-induced orbital eccentricity evolution over a continuous range of binary eccentricities spanning $0.0$ to $0.9$, and for the first time, we compute the corresponding binary semimajor-axis evolution rate needed to predict the fate of binaries at all eccentricities. Probing orbital element evolution over a continuous eccentricity range allows us to identify a steep transition in the disk and binary response that is responsible for the structure of attracting eccentricity solutions identified in \citet{Zrake+2020}.

\section{Numerical Methods}
\label{s:Numerical Setup}

Our set-up follows that of DD20 except that we vary the binary eccentricity at a fixed binary mass ratio of unity. We highlight the pertinent points below but refer the reader to DD20 and \citet{DuffellDISCO:2016} for more details.

We use the moving mesh code \disco~to solve the 2D equations of viscous, locally-isothermal hydrodynamics in the influence of the time changing gravitational potential of a binary on a fixed orbit. The relevant parameters setting the binary gravitational potential are the binary mass ratio $q \equiv M_2/M_1 \leq 1$; $M_1 + M_2 = M$, and the orbital eccentricity $0 \leq e < 1$. We choose units of $G=M=1$ and binary semimajor axis $a = 1$, so that one orbit of the binary is $2 \pi a^{3/2}/\sqrt{GM} \equiv 2 \pi \Omega^{-1} = 2 \pi$ in our units, for binary orbital frequency $\Omega$. 
Two remaining parameters characterize the isothermal fluid flow, the Mach number $\Mach$ and the coefficient of kinematic viscosity $\nu$. 

The locally-isothermal nature of the disk is enforced by setting the sound speed to a fixed function of the coordinates,
\begin{equation}
    c_s = \frac{\sqrt{\Phi_1 + \Phi_2}}{\Mach}
\end{equation}
where $\Phi_i$ is the gravitational potential of the $i^{\mathrm{th}}$ binary component. 
The gravitational potential is smoothed below the length scale $s = 0.5 a/\Mach = 0.05 a$,
\begin{equation}
    \Phi_j = \frac{G M_j}{\sqrt{ |\scr_{ij}|^2 + s^2 }},
\end{equation}
as required for a consistent representation of the 3D potential in the vertically averaged limit \citep{Mueller2DhydroGS+2012}. Here $|\scr_{ij}|$ is the distance from the $i^{\mathrm{th}}$ cell to the $j^{\mathrm{th}}$ binary component.

This prescription ensures that the sound speed approaches values expected for a constant Mach number and Keplerian fluid velocities throughout the disk. We choose a fiducial Mach number of $\Mach=10$. For a disk in vertical hydrodynamic equilibrium this corresponds to a disk aspect ratio of $H/r= \Mach^{-1} = 0.1$.

We choose a disk with a constant coefficient of kinematic viscosity $\nu = 10^{-3} a^2 \Omega$, different from $\nu$ of the standard $\alpha$-prescription. In terms of which, $\nu = \alpha/\Mach^2 a^2 \Omega$, and our choice of $\nu$ corresponds to $\alpha = 0.1$ at $r=a$. This results in a viscous time at $r=a$ of $\frac{2}{3} a^2/ \nu\approx 106 (2\pi \Omega^{-1}$).

We use a log grid in the radial coordinate  
with a fiducial choice of 512 radial cells and an outer boundary at $r_{\mathrm{out}}=50a$ providing a resolution of $\delta r \approx 0.016a$ at $r=a$ and ranging between $\delta r \approx 0.01 a - 0.02 a$ for the extremes of binary orbital radii simulated here.  We set the number of azimuthal cells at each radius to enforce equal cell aspect ratios.

The initial fluid variables constitute a uniform surface density $\Sigma_0$, a Keplerian orbital velocity for $r>a$ and rigid rotation with the binary for $r<a$. The radial velocity is initialized at the viscous drift rate $-\frac{3}{2}\nu/r$. 

Gas is removed from a region surrounding each binary component using the sink prescription of DD20.
The density of the $i^{\mathrm{th}}$ cell is removed as,
\begin{equation}
    S_\Sigma = - \gamma \Sigma \left( \mathrm{exp} \left( -\frac{ |\scr_{i1} |^4}{(2s)^4} \right) + \mathrm{exp} \left( -\frac{ | \scr_{i2} |^4}{(2s)^4} \right) \right),
\end{equation}
\noindent
for mass removal rate $\gamma$. We choose a fiducial value $\gamma = 1.0 \Omega$. The scale of the sink radius is chosen to be twice the gravitational smoothing length.

\begin{figure*}
\begin{center}$
\begin{array}{c}
\includegraphics[scale=0.9]{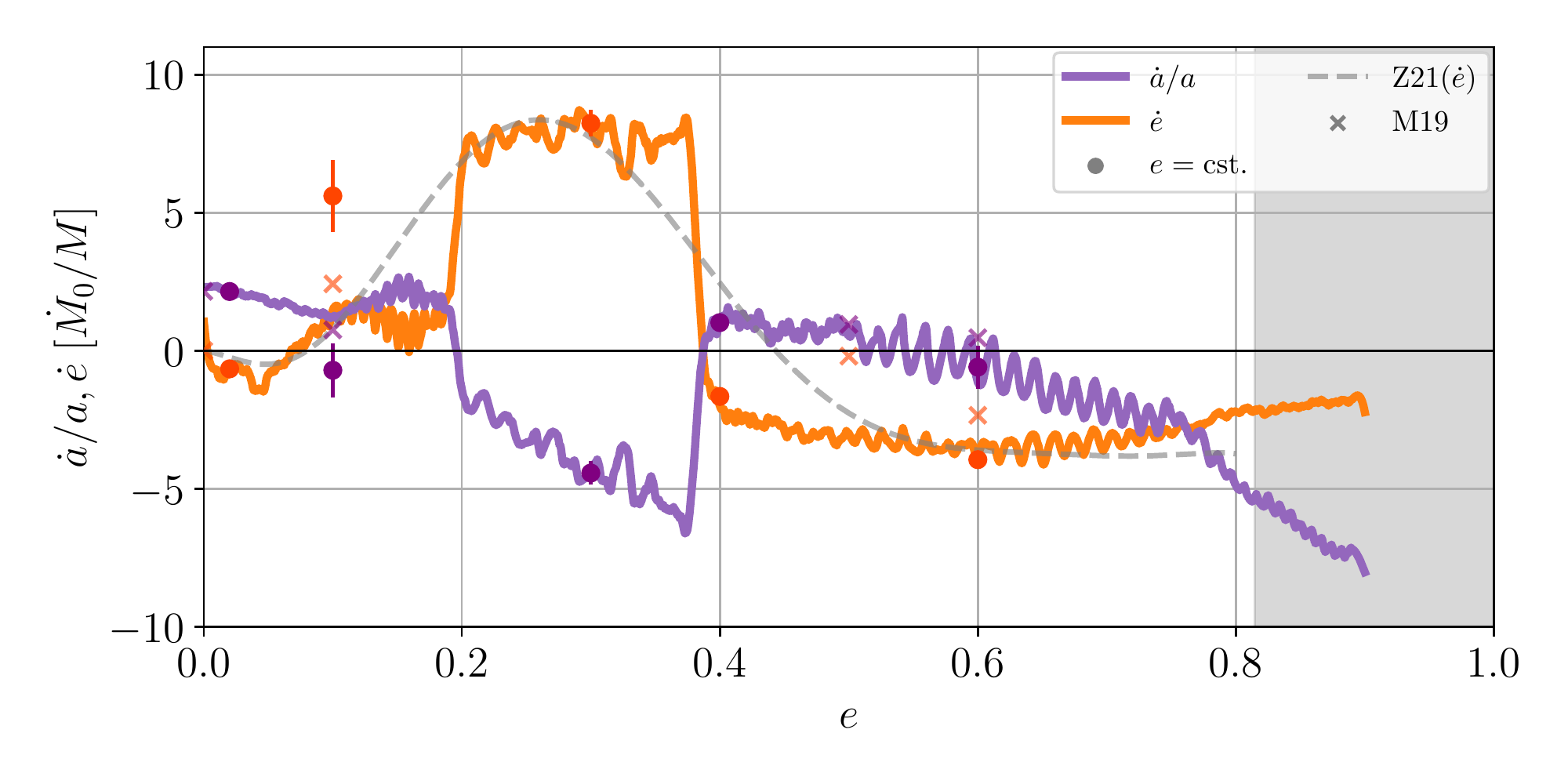}
\end{array}$
\end{center}
\vspace{-20pt}
\caption{
$\adot$ and $\edot$ (thick purple and orange lines) measured from our fiducial calculation overlaid with the results of constant eccentricity runs (circles). The fitting function for $\edot$ from \citet{Zrake+2020} (Z21) is plotted as the grey dashed line. The results from \citet{MunozLai+2019} (M19) are plotted as x's with the same $\adot$ (purple), $\edot$ (orange) color scheme. The grey shaded region is where twice the smoothing length becomes equal to the approximate size of the mini-disk at pericenter.
}
\label{Fig:aedot}
\end{figure*}

Our fiducial calculations fix the binary mass ratio at $q=1.0$ and vary the binary eccentricity. We first fix the orbital eccentricity to $e=0$ and run for $500$ orbits until we find an approximate steady state in the measured diagnostics discussed below. We use the output from this $e=0$ run to initialize a calculation that slowly grows the binary eccentricity from $e_0=0$ to $e_f$. We choose $e_f$ such that binary pericenter is always greater than two smoothing lengths, $a(1-e_f) = 2s$, resulting in $e_f=0.9$. We vary the binary eccentricity linearly in time,
\begin{equation}
    e_s(t) = e_0 + \frac{e_f}{2 \pi n_{\max}} t,
\end{equation}
for a total run lasting $n_{\max}$ binary orbits.
We choose $n_{\max}=2\times10^4$ based on agreement with constant eccentricity runs (see Figure \ref{Fig:aedot}), and a study that compared results\footnote{For larger $n_{\max}$ we find steeper transitions between disk states and a decreasing level of noise in the orbital evolution rates. Away from the disk transitions at $e\approx0.2$ and $e\approx0.4$, $\adot/a$ and $\edot$ have the same average value for all $n_{\max}$.} for $n_{\max}=5\times10^3$, $10^4$, and $2\times10^4$.
As the binary sweeps through this continuous range of eccentricities, we measure the rate of change of binary orbital semimajor axis $\adot(e)$ and eccentricity $\edot(e)$.

\subsection{Diagnostics}
The specific energy and angular momentum of the binary are
\begin{eqnarray}
\epsilon = - \frac{G M}{2 a},
\qquad l^2 = GM a (1-e^2). \nonumber
\label{Eqs:lilEL}
\end{eqnarray}
Differentiation with respect to time gives the binary evolution equations in terms of the specific torque $\ldot$ and power $\epsdot$ applied to the binary by the gas\footnote{Equivalent expressions written in terms of the total binary energy $E$ and angular momentum $L$, can be recovered by using that $L = \mu l$ and $E = \mu \epsilon$, for reduced mass $\mu$.},
\begin{eqnarray}
\frac{\adot}{a} &=& \frac{\Mdot}{M} - \frac{\epsdot}{\epsilon}, \\ \nonumber \\
\edot &=& \frac{1-e^2}{2e}\left[2 \frac{\Mdot}{M} -  \frac{\epsdot}{\epsilon} - 2\frac{\ldot}{l} \right].  
\label{Eq:aedot}
\end{eqnarray}

The accretion rate onto the binary, $\Mdot$, is measured using the above sink prescription. The specific power is computed by differentiating $\epsilon = \frac{1}{2}\mathbf{\dot{r}_b} \cdot \mathbf{\dot{r}_b} - GM/r_b$, using that $\mathbf{\ddot{r}_b}=\mathbf{f_g} - (GM/r^3_b)\mathbf{r_b}$,
\begin{equation}
\epsdot = \mathbf{v_b} \cdot \mathbf{f_g}  - \frac{G\Mdot}{r_b},
\end{equation}
for time-dependent binary separation $r_b$.
The specific torque is found from differentiating $l=\mathbf{r_b} \times \mathbf{\dot{r}_b}$,
\begin{equation}
\ldot = \mathbf{r_b} \times \mathbf{f_g}.
\end{equation}
The acceleration of the binary induced by the gas is measured directly as,
\begin{equation}
\mathbf{f_g} = \sum^{j=2}_{j=1}\sum_i dV_i\frac{G\Sigma_i}{|\scr_{ij}|^2}  \mathbf{\hat{ \hspace{-2pt} \scr}_{ij}} , 
\end{equation}
where the first sum is over each cell with volume $dV_i$ and gas-surface density $\Sigma_i$, and the second sum is over the binary components with $\mathbf{\hat{\hspace{-2pt} \scr}_{ij}}$ the unit vector pointing from binary component to cell. As the equations are scale free in the density, we normalize our results below in terms of the steady-state accretion rate at infinity $\dot{M}_0 = 3\pi \Sigma_0 \nu$, and the total binary mass $M$ \citep{MunozLai+2019}. We do not include a force caused by anisotropic accretion of momentum \citep[\eg, $\mathbf{f}_{\mathrm{acc}}$ in][]{MunozLai+2019} as we estimate its contribution to our main results to be minimal for the small sinks used here (see further discussion in \S~\ref{S:Discussion}).

To compute a time-averaged orbital evolution at a given value of $e$, and to arrive at the curves in Figure \ref{Fig:aedot}, we smooth the oscillating, raw output, which is computed 100 times per orbit\footnote{Increasing this to 1000 times per orbit does not affect our results.}. We compute $\adot$ and $\edot$ from this output via Eqs.~(\ref{Eq:aedot}) and then apply a Saviztky-Golay filter which employs a third-order polynomial fitting with smoothing wavelength chosen to be $\approx 444$ orbits, unless noted otherwise. This is chosen to approximately coincide with a viscous time at $r=2a$, the cavity precession time (\S \ref{s:Results1}), and $\Delta e = 0.02$.

For our fiducial calculations, where we grow the eccentricity linearly in time, we apply this smoothing across the time series after the initial $500$ orbits where $e=0$. For the constant eccentricity runs, we allow $\adot$ and $\edot$ to reach a quasi-steady state, and apply the \textit{same} smoothing to the final $\sim1000$ orbits of these runs. The mean and standard deviation of the smoothed output are taken to build the scatter points and error bars in Figure \ref{Fig:aedot}.

\section{Results}
\subsection{Orbital semimajor Axis and Eccentricity Evolution}
\label{s:Results1}

Figure \ref{Fig:aedot} displays our primary result, the orbital semimajor-axis and eccentricity evolution of the binary as a function of eccentricity. We plot $\adot(e)$ in purple and $\edot(e)$ in orange. Also plotted in Figure \ref{Fig:aedot} are results of constant eccentricity runs for $e=\left\{0.02, 0.1, 0.3, 0.4, 0.6\right\}$. Figure \ref{Fig:aedot} also compiles results of the most directly comparable studies, the fitting function for $\edot$ from \citet{Zrake+2020} (dashed grey line) and results from Table 1 of \citet{MunozLai+2019}.  The grey-shaded region delineates where the gravitational smoothing length is larger than the mini-disk size, approximated by adapting the numerical fit for circular orbits from \citet{Roedig+2014}, $r_d \approx 0.27 a(1-e) q^{0.3}$. Here forces exerted by gas in the mini-disks are likely less accurate.

Because both $\adot$ and $\edot$ depend only on $e$, it is the orange $\edot$ curve that dictates the binary's time evolution. The quantity $\edot$ is negative for $e \lesssim 0.07$ and $e \gtrsim 0.39$ and positive inbetween, and nearly constant for $0.2\leq e \leq 0.39$. A defining feature of both curves is a steep change in behavior near $e=0.2$ and $e=0.4$. The first of which results in a change in sign of $\adot$ at $e = 0.2$ and an increase in $\edot$ by a factor of $8$, the second of which results in the rapid change in sign of both $\edot$ and $\adot$ causing both to become zero at $e\approx0.39$. Note how the continuous eccentricity approach reveals the steepness of the $e=0.2, 0.4$ transitions.

\begin{figure*}
\begin{center}$
\begin{array}{c}
\hspace{-10pt}
\includegraphics[scale=0.27]{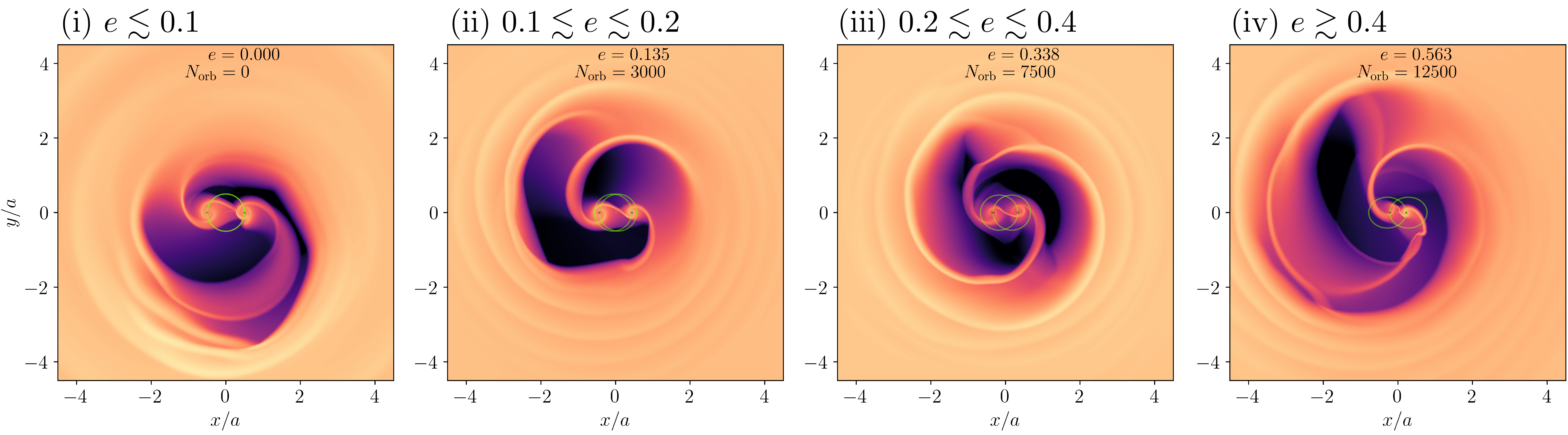} \hspace{-15pt}
\end{array}$
\end{center}
\vspace{-20pt}
\caption{
Log-surface-density snapshots at pericenter for the fiducial calculation which linearly increases orbital eccentricity from $e=0$ to $e=0.9$ over $2 \times 10^4$ binary orbits. We display a representative snapshot from each of the regimes (i)-(iv) discussed in \S \ref{s:Results1}. The binary components and their orbital tracks are plotted in green.}
\label{Fig:dens_eccs}
\end{figure*}

To elucidate the cause of this behavior, Figure \ref{Fig:dens_eccs} displays snapshots of log-gas-surface-density when the binary is at pericenter for binary eccentricities representative of four regimes separating distinct disk responses:

\paragraph{(i) Circular Binary, Lopsided Disk ($e\lesssim 0.1$)} The left panel of Figure \ref{Fig:dens_eccs} shows the lopsided disk structure reported in many works that consider $e=0, q=1$ binaries \citep[\eg,][to name a small subset]{MM08, DHM:2013, ShiKrolik:2012}. This constitutes an elongated cavity that precesses on a much longer timescale than the orbital period and is punctuated by an overdensity that orbits the cavity edge once every $\sim 5$ orbits \citep[\eg,][]{PG1302MNRAS:2015a}. 

\paragraph{(ii) Mildly Eccentric Binary, Asymmetric Disk ($0.1 \lesssim e \lesssim 0.2$)} In the second panel of Figure \ref{Fig:dens_eccs}, the overdensity and corresponding 
$\sim 5$ orbit accretion-rate periodicity disappears, but the elongated, precessing cavity remains.

\paragraph{(iii) Eccentric Binary, Origin-Symmetric Disk ($0.2 \lesssim e \lesssim 0.4$)} When the binary reaches an eccentricity of $e\sim0.2$, the cavity elongation is diminished giving rise to a disk with symmetry about the origin $\Sigma(x,y) \rightarrow \Sigma(-x,-y)$; the third panel in Figure \ref{Fig:dens_eccs} shows nearly equal strength streams reaching the binary from both sides of the cavity.

\paragraph{(iv) Highly Eccentric Binary, Asymmetric Disk ($e\gtrsim0.4$)} 
Above $e\sim0.4$, the disk again becomes elongated and precesses slowly around the binary. At these larger eccentricities (especially for $e\gtrsim0.5$), the density structure in the cavity is generally more complex due to the eccentric binary spanning a larger range of separations over the course of its orbit as it pulls in gas streams and propels them back out to shock into the surrounding disk \citep[see also][]{Moesta+2019}.

\begin{figure}
\begin{center}$
\begin{array}{c}
\hspace{-15pt}
\includegraphics[scale=0.515]{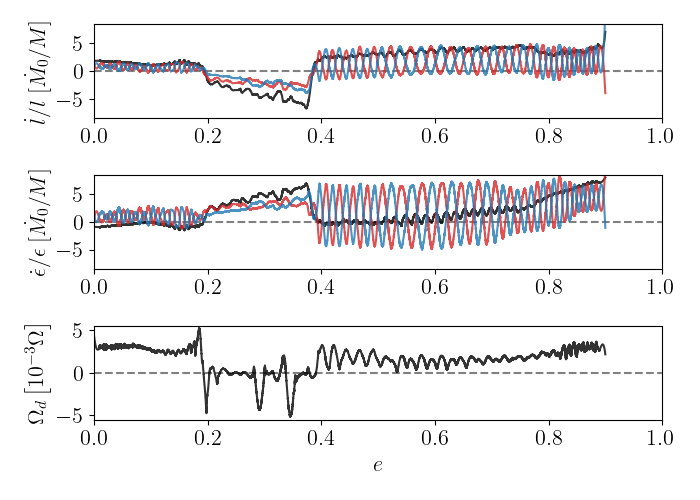}
\end{array}$
\end{center}
\vspace{-20pt}
\caption{
{\em Top two panels:} specific torque and power vs. $e$ for each binary component (red and blue), and total (black). Trading between binary components due to disk precession is well resolved over the eccentricity sweep. Direct association of disk precession with these oscillations is shown in the {\em bottom panel}, which plots the disk precession rate in units proportional to the binary orbital frequency and highlights that disk precession halts for $0.2\lesssim e \lesssim 0.4$. 
}
\label{Fig:TPs}
\end{figure}

Delineation into these regimes is further supported by Figure \ref{Fig:TPs}. The top two panels of Figure \ref{Fig:TPs} show the smoothed specific torque per specific binary angular momentum and the specific power per specific binary energy. Each panel shows contributions from each binary component in red and blue with the total in black. In the first two regimes, for $e\lesssim 0.2$, and in the high eccentricity regime, for $e\gtrsim0.4$, the precession of the cavity can be seen in the completely out of phase oscillations of the (red and blue) component-wise quantities.

Disk precession causes these oscillations because, relative to the binary center of mass, there is a near side and a far side to the asymmetric disk structure 
(see all but panel three of Figure \ref{Fig:dens_eccs}). 
For an equal-mass binary on a circular orbit, each binary component has the same interaction with the disk, just half of an orbit out of phase with the other. However, an eccentric orbit breaks this symmetry by differentiating which binary component interacts with the near (far) side of the disk at apocenter (pericenter) or vice versa \citep[see also][]{Dunhill+2015, MunozLai+2016}. For example, at apocenter, one binary component plunges into the disk's near side while the other stays far from the disk's far side. After the disk precesses by a one-half rotation, the situation is reversed. We run our fiducial eccentricity-varying calculation for long enough to smooth over these greater-than-orbital-timescale variations.

The bottom panel of Figure \ref{Fig:TPs} plots the disk precession rate $\Omega_d$ by computing the time derivative of the phase of the domain-integrated complex quantity $\int^{r_{\mathrm{out}}}_0 \int^{2\pi}_0{ \Sigma(t, r,\phi) \mathrm{e}^{i \phi} r dr d\phi}$.
For $e \sim  0.0$ the disk precession period is $\approx 350 (2\pi\Omega^{-1})$. This drops to $\approx 300 (2\pi\Omega^{-1})$ at $e\approx 0.07$ before rising to $\approx 400 (2\pi\Omega^{-1})$ at $e\approx 0.1$, coincident with disappearance of the lump and the onset of regime (ii). The rate remains steady over $0.1 \lesssim e \lesssim 0.17$ until a steep doubling of the precession rate between $e\approx0.17-0.19$, and a halt in precession for $0.2 \lesssim e \lesssim 0.4$ (with the exception of two excursions at $e=0.291$ and $e=0.345$). Zero precession in this regime is indicative of the origin-symmetry of the disk (third panel of Figure \ref{Fig:dens_eccs}), for which precession about the origin is not possible. Cavity precession resumes for $e\gtrsim0.4$ but at approximately half the rate observed in regime (i), until rising again for $e\gtrsim0.8$, where the relative size of the gravitational softening length causes results to become suspect.

Finally, notice that the constant eccentricity runs agree very well with the continuous sweep, except for at $e=0.1$. Here, density snapshots from the constant eccentricity run display similarities to the origin-symmetric state, in disagreement with Figure \ref{Fig:dens_eccs}. To understand the origin of this discrepancy, we run another $e=0.1$ run, but starting from steady-state, asymmetric $e=0$ initial conditions. The result of this run matches the result of the sweep calculations. After $\approx 3000$ orbits, both $e=0.1$ runs hold steady in their respective states. So there is either an inherent initial state memory in this eccentricity range, or the transition time to the origin-symmetric state at $e\sim0.1$ takes longer than a few thousand orbits. Why this isn't an issue for the $e=0.4$ transition may be due to the shorter transition time from the origin-symmetric to the asymmetric state discussed in \S~\ref{s:Consequences for Binary Evolution}. For rapidly evolving systems, such direction-dependent evolution may be physical.

\subsection{Consequences for Binary Evolution}
\label{s:Consequences for Binary Evolution}
When the disk is much less massive than the binary, the solutions for $\adot$ and $\edot$ provide all that is needed to determine the long-term evolution of an equal-mass binary interacting with our fiducial gas disk. The possible solution behaviors for $a(t)$ and $e(t)$ can be greatly simplified by analyzing the shape of our measured functions for $\adot$ and $\edot$.

The solution for $\edot(e)$ has three zeros. Those at $e=0$ and $e\approx0.4$ are attractors, and one at $e\approx0.1$ is a repulsive point that acts as the divide between the $e=0$ and $e\approx0.4$ attractors. Hence, the behavior of $a(t)$ and $e(t)$ near $e=0$ and $e\approx0.4$ is the most important to understand. The attractor at $e=0$ attracts binaries with $e\lesssim0.1$. Since $\adot(e=0)>0$ in Figure \ref{Fig:aedot}, binaries with $e\lesssim0.1$ are destined to expand on circular orbits.

The other attractor at $e\approx0.4$, which we denote as $e_*$ deserves further investigation. The same disk transition that causes the change in sign of $\edot$ also causes $\adot$ to change sign at the same value of $e=e_*$. At first glance this implies that all binaries with $e\gtrsim 0.1$ are destined to evolve toward orbits with $e = e_*$ and an unchanging semimajor axis, or that fine tuning of the exact value of the zeros of $\adot$ and $\edot$ would muddy predictions for orbital evolution. However, the steep change in binary response at $e=e_*$ is caused by the disk state change discussed in \S \ref{s:Results1}, and this occurs over a finite-transition lag time.

\begin{figure*}
\begin{center}$
\begin{array}{ccc}
\hspace{-10pt}
\includegraphics[scale=0.35]{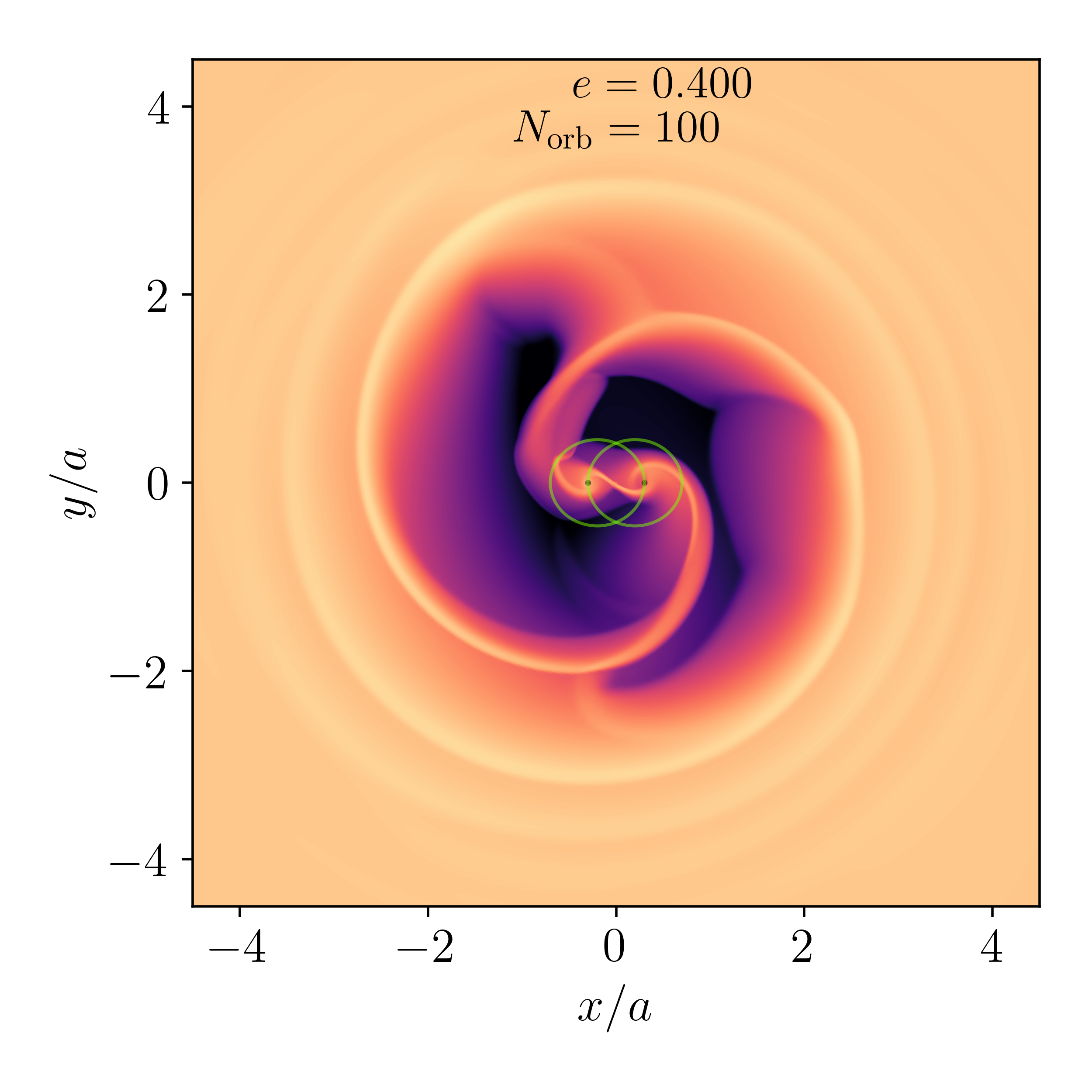}&
\hspace{-10pt}
\includegraphics[scale=0.45]{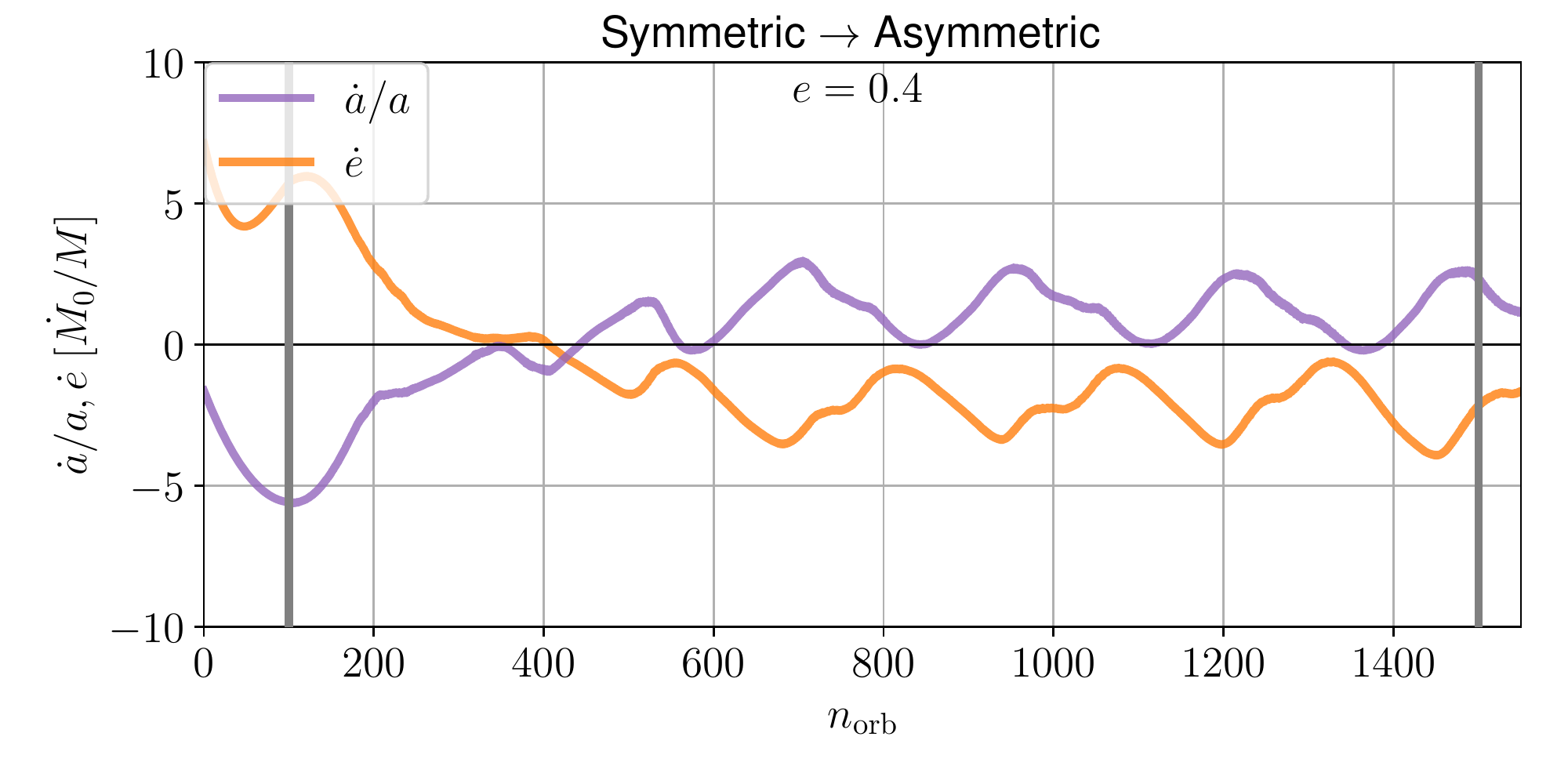}  &
\hspace{-10pt}
\includegraphics[scale=0.35]{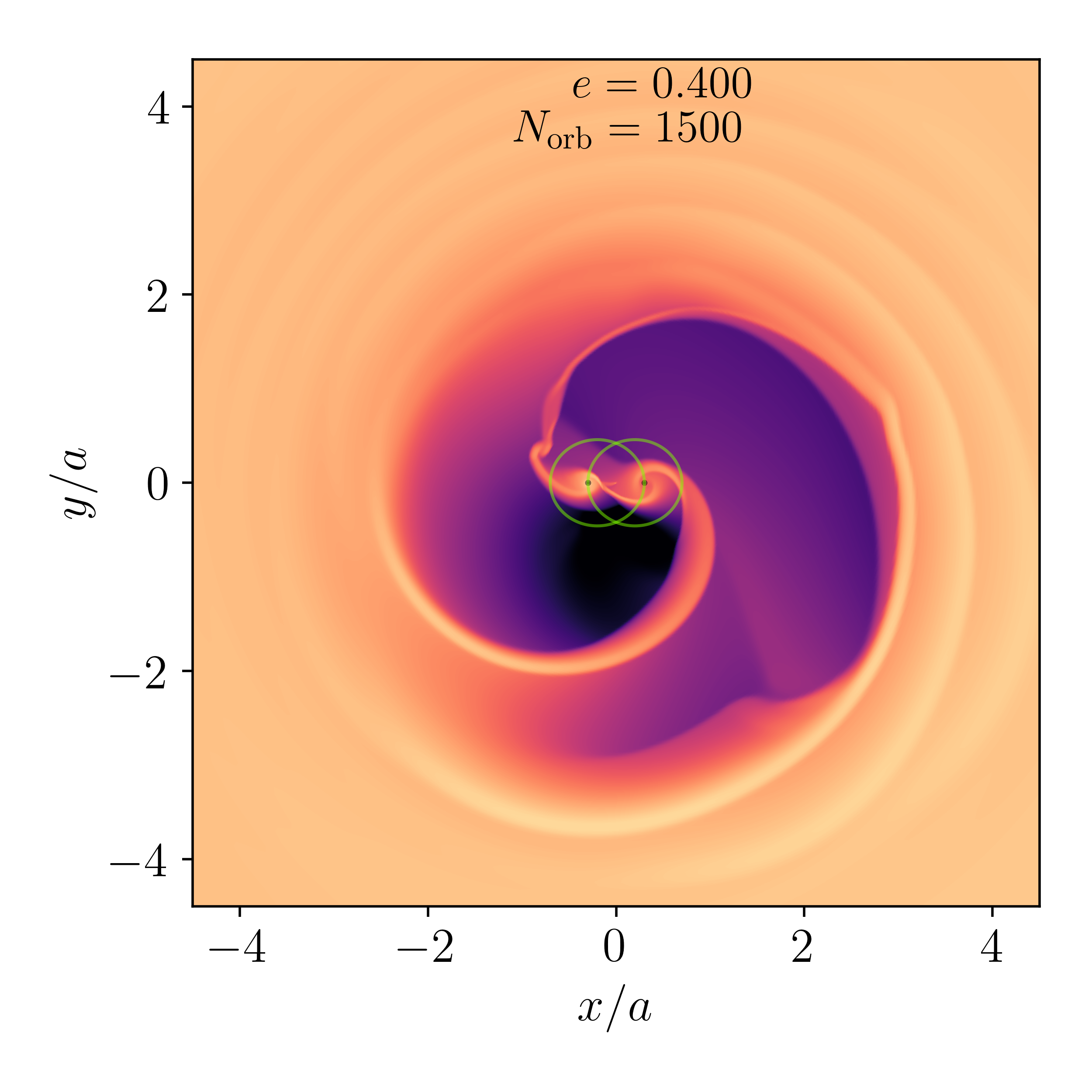} \\
\hspace{-10pt}
\hspace{-10pt}
\includegraphics[scale=0.35]{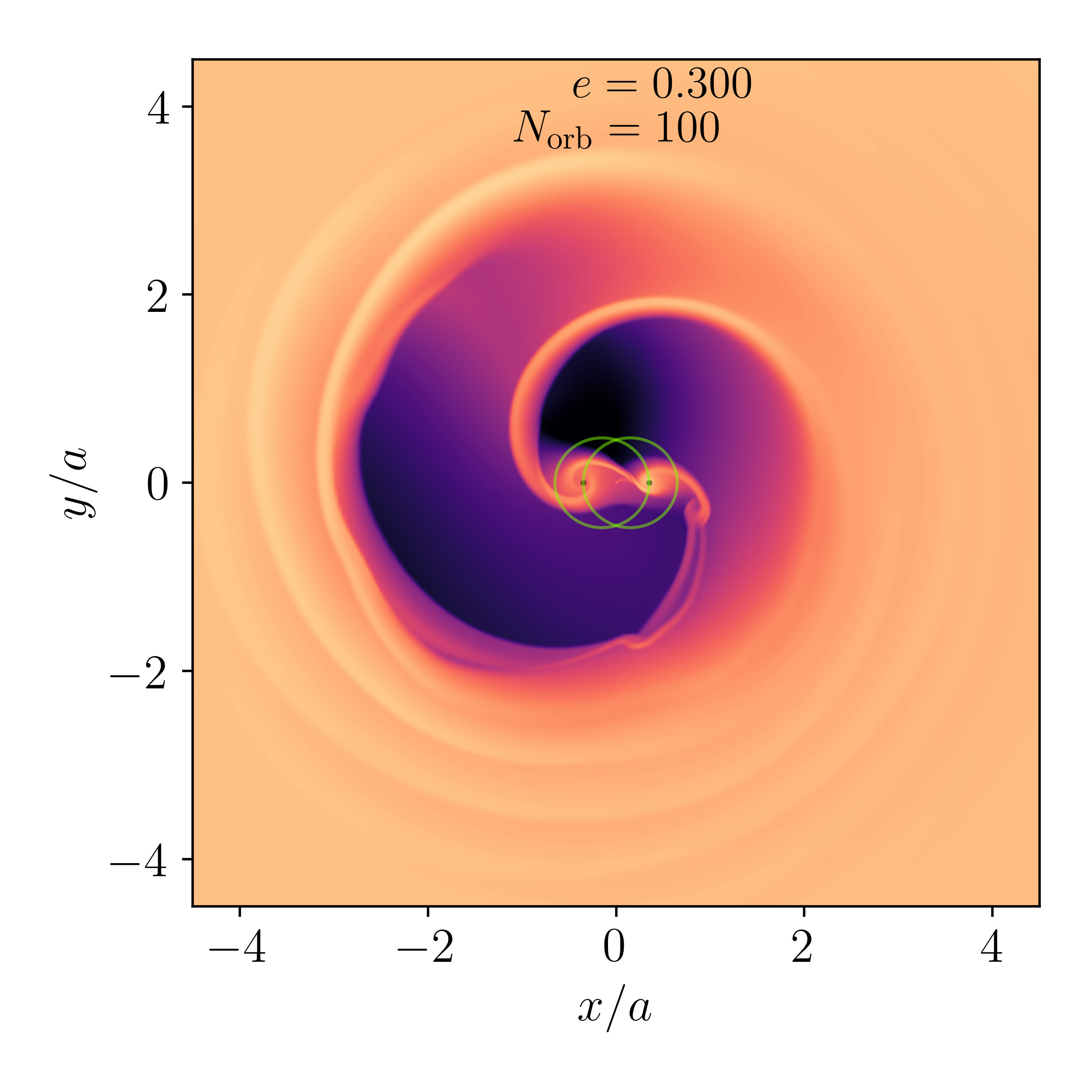}&
\hspace{-10pt}
\includegraphics[scale=0.45]{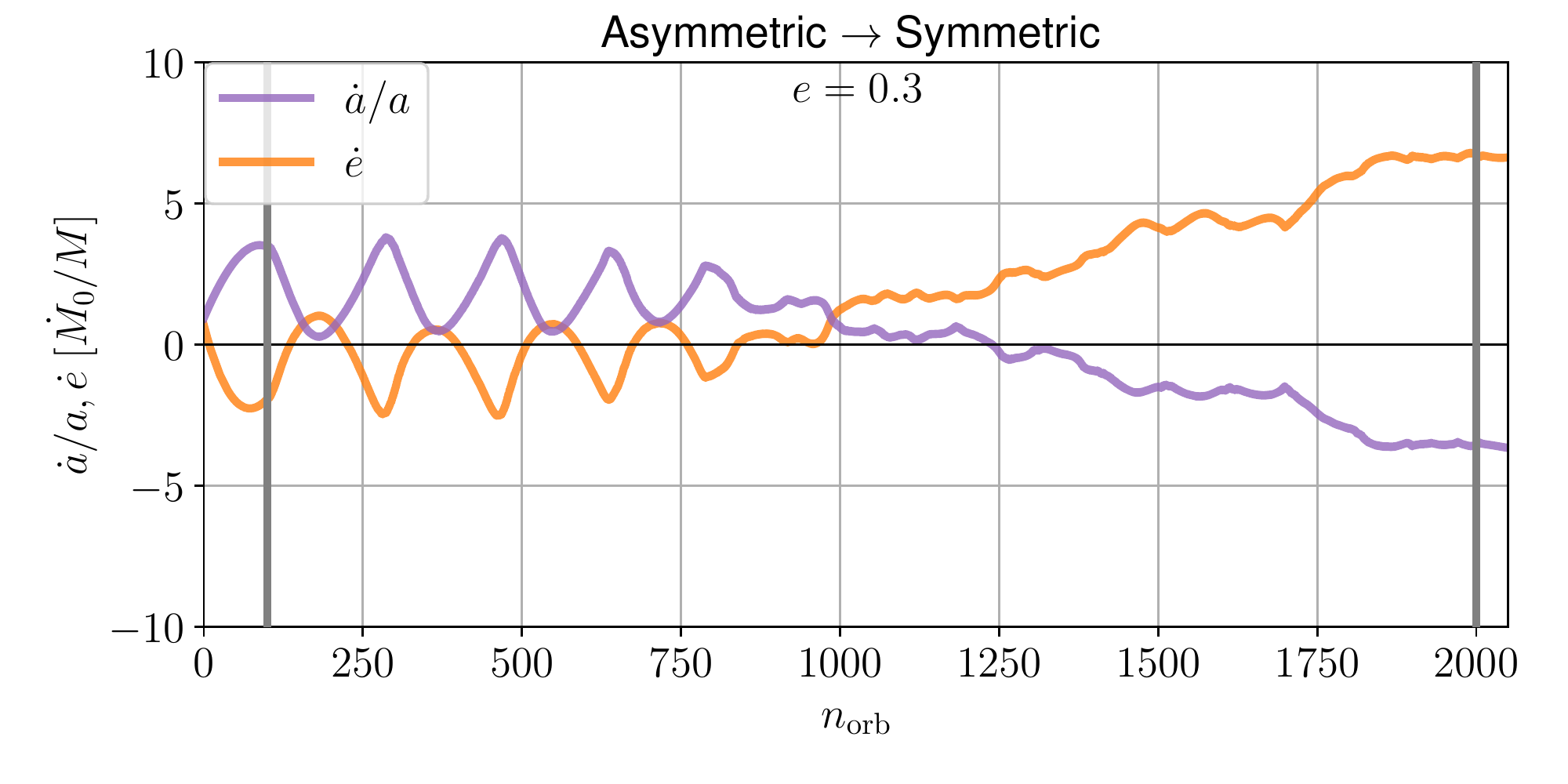} &
\hspace{-10pt}
\includegraphics[scale=0.35]{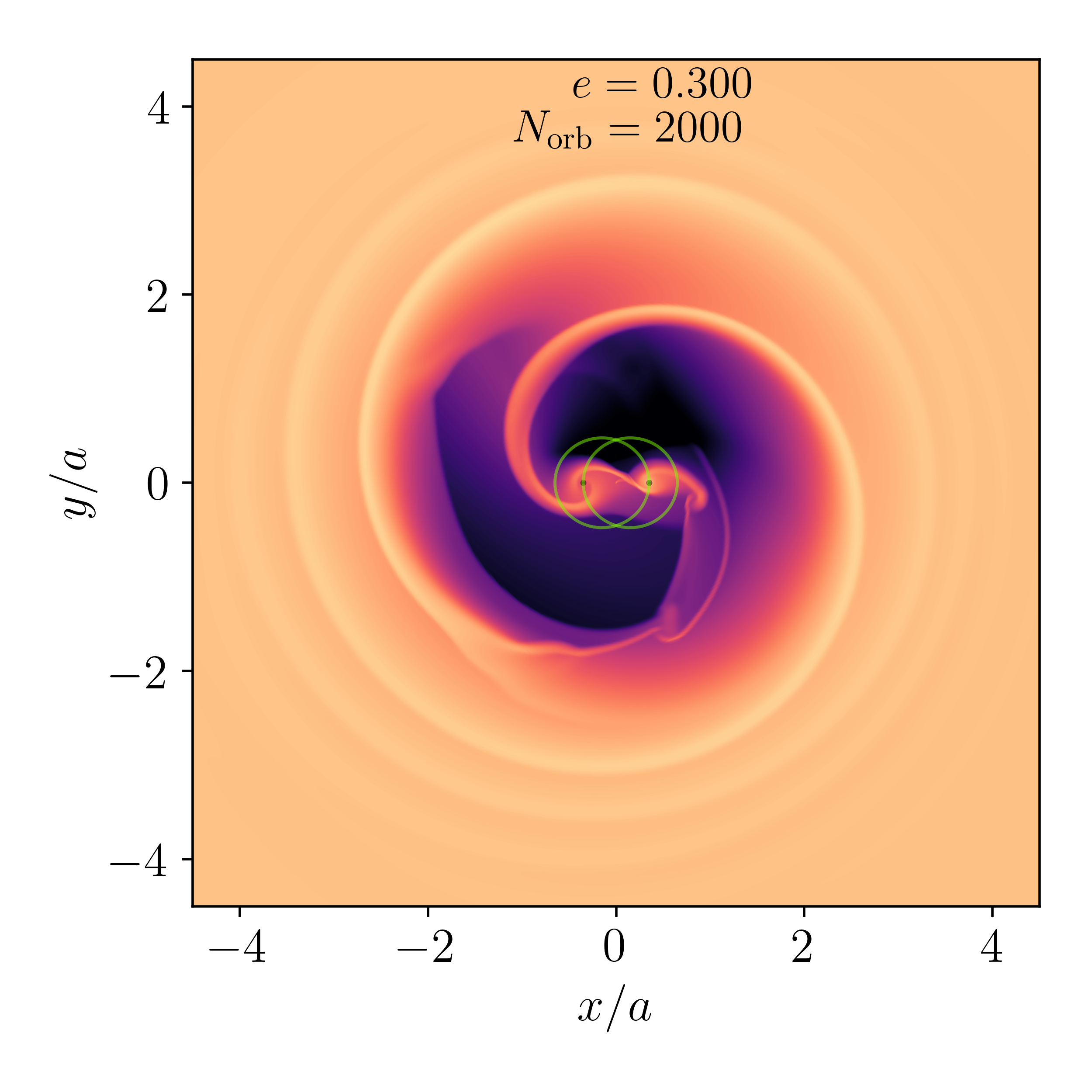}
\hspace{-10pt}
\end{array}$ 
\end{center}
\vspace{-10pt}
\caption{
The time evolution of $\adot$ and $\edot$ during transitions to and from the origin-symmetric state and the asymmetric, precessing state. Log-surface-density snapshots are taken at times indicated by the vertical grey lines. 
{\em Top:} for fixed binary eccentricity $e=0.4$, starting from the quasi-steady state output of a $e=0.3$ run. The disk and binary response transition after $\sim 700$ binary orbits. 
{\em Bottom:} for a fixed binary eccentricity of $e=0.3$, starting from the quasi-steady state output of a $e=0.4$ run. The disk and binary response transition after $\sim 1900$ binary orbits. Smoothing is carried out over $\approx 200$ orbits to show oscillations due to disk precession.
}
\label{Fig:trans_lag}
\end{figure*}

This lag time is explored in Figure \ref{Fig:trans_lag} where we use the output of the constant eccentricity runs at $e=0.3$ (origin-symmetric state) and $e=0.4$ (asymmetric state) as the input for new constant eccentricity calculations but now on the other side of the transition eccentricity. That is, the output of the $e=0.3$ ($e=0.4$) run is the input for a new $e=0.4$ ($e=0.3$) run. Figure \ref{Fig:trans_lag} shows that the transition does indeed occur in both directions and with a lag time of $\sim 700$ binary orbits going from the $e=0.3$ initial conditions to the $e=0.4$ steady state (origin-symmetric to asymmetric), and $\sim 1900$ binary orbits going in the opposite direction (asymmetric to origin-symmetric).

This lag between origin-symmetric and asymmetric, precessing disk states will cause a continual overshooting across $e_*$ resulting in an oscillation of the binary eccentricity. Because $\edot$ and $\adot$ are asymmetric around $e_*$, this can cause a net drift of the binary semimajor axis.

To determine the nature of the semimajor-axis drift, which a binary on a nonfixed orbit would experience, we note that the behavior of $\adot$ and $\edot$ around $e_*$ resembles a steep transition connecting constant rates. Hence, we model the behavior around the attractor $e_*$ as such, with the set of ordinary differential equations,
\begin{eqnarray}
\dot{a} &=& \adot_l \Hside\left[ e_* - e\left(t-\tau_{r,l} \right)  \right] + \adot_r \Hside\left[ e\left(t-\tau_{r,l} \right) - e_*\right]  \nonumber  \\
\dot{e} &=& \edot_l \Hside\left[ e_* - e\left(t-\tau_{r,l} \right) \right] + \edot_r \Hside\left[ e\left(t-\tau_{r,l} \right) - e_*\right],
\label{Eq:ToyModDEq}
\end{eqnarray}
where $\left\{\adot_l, \adot_r, \edot_l, \edot_r\right\}$ are the constant rates on either side of $e_*$ and $\Hside$ is the unit step function. We include the finite-transition time between disk states with the introduction of $\tau_{r,l}$, which is the transition time from origin-symmetric to asymmetric states, $\tau_r$, if $\edot>0$ and the reverse, $\tau_l$, otherwise.

Before solving these equations, we find an analytical solution for the average rate of change of the semimajor axis $\left< \adot_* \right>$ by realizing that, within our simplified model, the behavior of the binary semimajor axis near $e_*$ is simply an asymmetric sawtooth in time characterized by the slope at which it rises $\adot_r$, the slope at which it decays $\adot_l$, and the time spent in each regime, which is set by the quantity $\xi \equiv |\edot_l/\edot_r|$, $\tau_r$, and $\tau_l$. 
Then at the transition eccentricity $e_*$, the average rate of change of the binary orbital parameters is,
\begin{equation}
    \left< \adot_* \right> = \frac{\adot_l + \adot_r \xi }{1 + \xi },
    \qquad 
     \left< \edot_* \right> = 0,
    \label{Eq:ToyModSoln}
\end{equation}
which remarkably does not depend on the transition times between origin-symmetric and asymmetric states, except that both are nonzero. This arises because the change in $a$ over one oscillation cycle around $e_*$ and the total duration of a cycle depend identically on the transition times $\tau_{r,l}$. The delay times do determine the average of the  eccentricity oscillations, $\left< e \right> = e_* + 0.5\left(\edot_l \tau_r + \edot_r \tau_l \right)$.

\begin{figure}
\begin{center}$
\begin{array}{cc}
\includegraphics[scale=0.37]{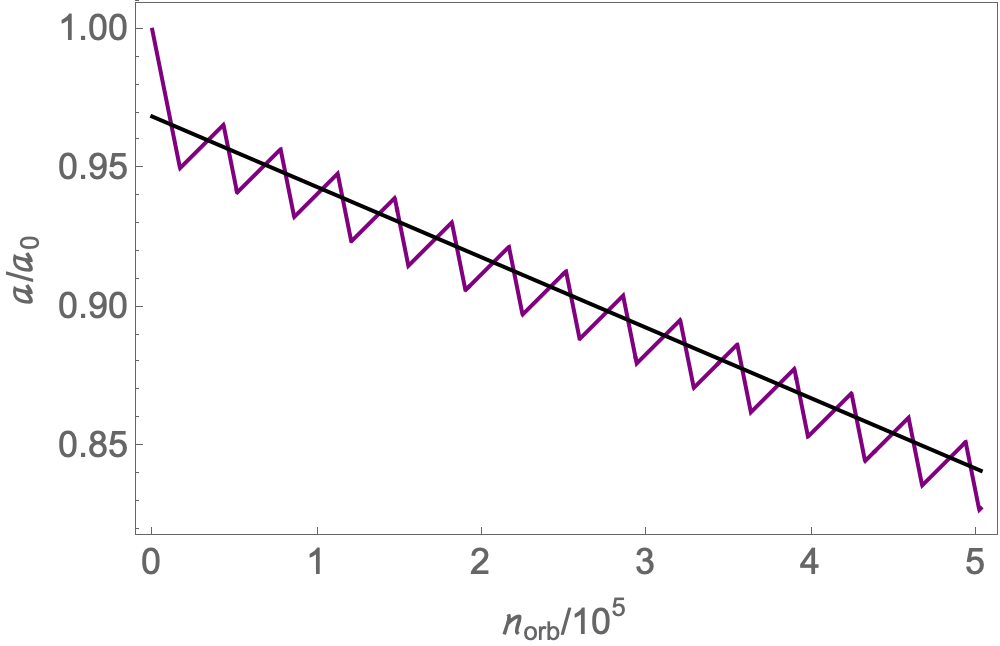} \vspace{-14pt} \\
\hspace{-5pt}
\includegraphics[scale=0.382]{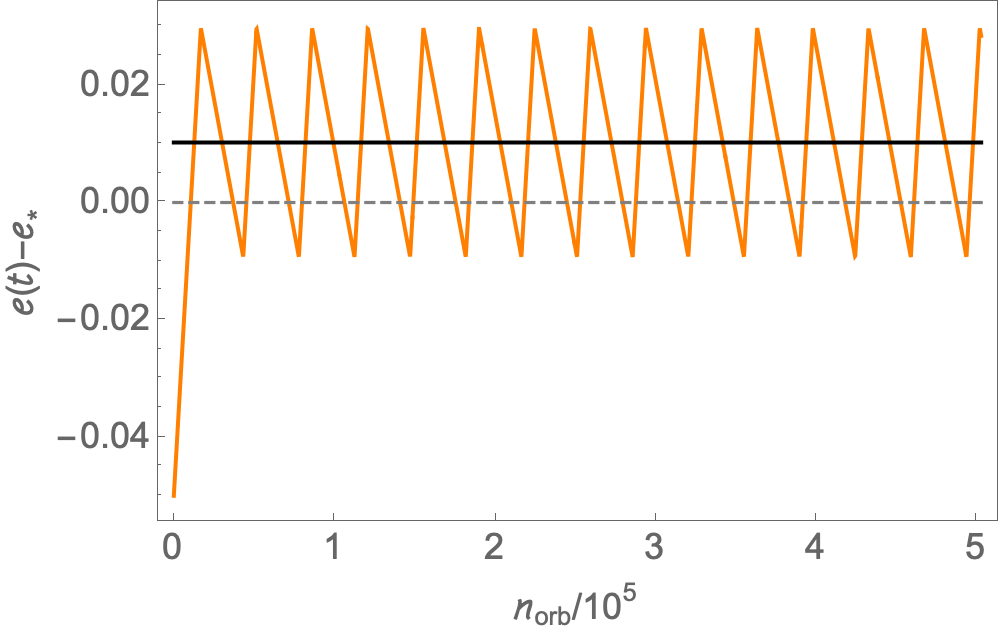} 
\end{array}$
\end{center}
\vspace{-15pt}
\caption{
Binary orbital evolution near the transition eccentricity $e_*$, as modeled by Eqs. (\ref{Eq:ToyModDEq}). The black lines are the analytic average expectations given by integrating Eqs. (\ref{Eq:ToyModSoln}). 
We choose a density scaling of $\Sigma_0 a^2/M=10^{-5}$ to set the x-axis scale. Note that $\left<e\right>$ can differ from $e_*$, marked by the dashed grey line.
}
\label{Fig:ToyMod}
\end{figure}

The condition for a decaying binary semimajor axis is,
\begin{equation}
   \left|\frac{\adot_l}{\adot_r}\right| \geq \left|\frac{\edot_l}{\edot_r}\right| .
\end{equation}
Using approximate measured values from Figure \ref{Fig:aedot}, $\edot_l=8$, $\edot_r=-2.5$, $\adot_l=-5$, $\adot_r=1$, we find,
\begin{equation}
    \adot_* \approx -0.43 a \left[\dot{M}_0/M \right]
\end{equation}
for initial semimajor axis $a$. For Eddington accretion rates, $a/\adot$ corresponds to $2.33$ Eddington times. 
The same result will arise whether or not the binary first approaches $e_*$ from the left or right.
Figure \ref{Fig:ToyMod} plots the solutions to Eqs. (\ref{Eq:ToyModDEq}) for $a(t)$, and $e(t)$, and their analytic averages, assuming $\tau_r=\tau_l=1000 (2 \pi \Omega^{-1})$ for simplicity.

\section{Discussion}
\label{S:Discussion}
For thin, locally-isothermal circumbinary disks around equal-mass binaries, and for a fiducial set of disk parameters, we have shown that the binary orbital evolution is driven to two attracting solutions: (i) expanding circular orbits for initial eccentricities $e_0\lesssim 0.1$, and (ii) decaying orbits with orbital eccentricity oscillating around $e \approx 0.4$ for initial eccentricities $e_0\gtrsim 0.1$. 
Importantly, we have linked this behavior to a transition in the disk response that results in an origin-symmetric disk state for $0.2 \lesssim e \lesssim 0.4$. Hence, the robustness of these relatively simple results can be vetted by better understanding this disk transition. 

\citet{MirandaLai+2017} find a similar disk transition in their 2D isothermal hydrodynamical calculations, which sample a few different eccentricities and use the same fiducial disk parameters but cut out the region of the domain containing the binary \citep[see also][]{Thun+2017}. They attribute the existence of precessing asymmetric and nonprecessing states to eccentricity excitation at eccentric Linblad resonances \citep[ELRs;][]{Lubow:1991} competing with viscous damping. If this is the case, then future analytical work can predict the change in onset of the origin-symmetric state for different disk viscosity and Mach number. For example, \citet{Tiede+2020} and \citet{HeathNixon:2020} show that the expansion of $e=0, q=1$ binary orbits can reverse for higher Mach number disks. Hence, the robustness of our results should be understood in light of resonant theory and numerical calculations like those presented here, but for different disk Mach numbers and viscosities.

Further, we have identified the disk precession rate, which encodes disk symmetry about the origin as a key diagnostic of the disk and eccentric binary response. Future work should also carry out a quantitative investigation of \textit{disk} eccentricity in our scenario, as studied in many previous works \citep[\eg,][]{GoodchildOgilvie:2006, MunozLithwick:2020}.

Our results are in good agreement with comparable studies for $q=1$ binaries (using different codes). 
Our initial $e=0, q=1$ calculation from which we start the eccentricity sweep finds, in units of $\Mdot_0 a^2 \Omega$, $\dot{L} \approx 0.8$ after 500 orbits\footnote{Corresponding to $\dot{l}/l = 1.9-1.6 [\dot{M}_0/M]$.}, and $\dot{L} \approx 0.7$ after 5000 orbits, matching the range found in \cite{MirandaLai+2017} ($\dot{L}\approx0.8$), \cite{MunozLai+2019} ($\dot{L}\approx0.676$), \cite{MoodyStone:2019} ($\dot{L}\approx0.723$), \citep{Tiede+2020} ($\dot{L}\approx0.79$)\footnote{Using a $\sqrt{2}$ larger coefficient of kinematic viscosity.}, and DD20 ($\dot{L}\approx0.6$) for the same disk and binary parameters.  
Figure \ref{Fig:aedot} shows agreement with the $\edot$ measurement of \citet{Zrake+2020} except for the steepness of the transitions that we uniquely probe using a continuous eccentricity sweep. It also shows that \citet{MunozLai+2019}, while not sampling between $e=0.2-0.4$, do find a similar trend at high eccentricity, though with offset values of $\edot$. While it is not clear what causes these smaller differences at nonzero eccentricity, we note that the main physical difference between this study and the other two is the use of a constant viscosity (as opposed to an $\alpha$-viscosity) prescription, though this cannot explain the differences between \citet{Zrake+2020} and \citet{MunozLai+2019}. Otherwise, the three codes employ different numerical approaches that could be explicitly contrasted in future code-comparisons.

Another possibility for differences in calculated rates could come from the treatment of accreted momentum and the sink prescription. \citet{DittRyan:2021} implement an improved sink prescription following \cite{Dempsey+2020} which enforces a zero-torque boundary at the sink radius by conserving the angular momentum about each point mass when removing gas. This prevents artificial density depletion near the sink and enforces zero `spin torque' \citep[\eg,][]{MunozLai+2019} applied to the binary component.
Comparison of runs with and without this sink prescription provide an estimate for the importance of such torques and dependence on the sink prescription, which may also affect anisotropic accretion forces. We tested a version of this momentum conserving sink prescription (using shorter, $n_{\max}=500(2\pi\Omega^{-1})$ runs) and find that it does not change our main results. It does introduce small quantitative changes in $\adot$ and $\edot$ at the $e\approx0.2$ and $e\approx0.4$ disk transitions, and becomes increasingly important for $\adot$ at $e\geq 0.7$. Future work will further explore this effect.

However, as DD20 shows, results vary greatly with mass ratio; drastic changes in the disk and orbital response arise for $q\lesssim 0.05$, owing to a  mass-ratio-induced disk transition \citep{D'Orazio:CBDTrans:2016}. While future work should aim to understand the disk response for different mass ratios, we note that the long-term behavior of the binary will trend to equal-mass binaries (DD20), and hence our results are relevant for a wider range of initial binary mass ratios than just $q=1$.

When the local disk is much less massive than the binary, $\adot$ and $\edot$ are functions only of $e$, and our treatment of the orbital evolution in Eqs. (\ref{Eq:ToyModDEq}) is valid. Rapid changes in $\edot$ and $\adot$ caused by a very massive disk could introduce dependence on $a$, or induce binary apsidal precession that affects disk-binary apsidal locking. Hence, behavior at this critical eccentricity, modeled here with simplified evolution equations should be investigated further in light of these possibilities, eventually with a live binary.

We assume an infinite gas reservoir. Recent works show that while finite disks never reach a formal steady state, they do supply torques and mass accretion rates in agreement with infinite disks \citep{Munoz_FinDsk+2020, Tiede+2020}. Hence the attracting solutions presented here will be relevant for a sufficiently long-lasting disk supply. Otherwise, Figure \ref{Fig:aedot} provides the interim solutions for binary orbital evolution during the bulk of the disk lifetime.

The continuous range of binary eccentricities explored here provide a unique data set for probing binary accretion rates as a function of eccentricity, as was done for the binary mass ratio \citep[][DD20]{DHM:2013, Farris+2014}, and also for measuring disk-induced binary apsidal precession rates, which could have important dynamical consequences for accreting, compact-object binaries. Both are the subject of forthcoming work.

This disk transition may be important for the evolution of misaligned, eccentric-binary+disk systems \citep[\eg,][]{Nixon+2013, AlyNixon+2015, MoodyStone:2019}, which could be investigated in future 3D studies. There may be implications for observed, misaligned stellar binary-disk systems \citep[\eg,][]{JensenAkeson:2014}, the spins of the binary components \citep{Gerosa_spinalign+2015}, and the Kozai-Lidov mechanism for accreting systems \citep[\eg,][]{Smallwood+2021}.

\section{Conclusion}
For a continuous range of eccentricities spanning to $0.9$, we have calculated the coupled evolution of binary semimajor axis and eccentricity due to interaction with a thin, locally-isothermal circumbinary disk. We find that two attractor solutions for binary eccentricity distill the long-term behavior into two cases: (i) circular, expanding orbits, and (ii) binaries with decaying semimajor axes with eccentricity oscillating around a critical value near $e=0.4$. The nature of the latter solution is set by a physical transition in the disk. Hence, this work offers not only a simple prescription for long-term, eccentric binary+disk evolution, but importantly offers a physical description. This disk state-change must be investigated further to vet its robustness to a range of disk and binary parameters and so further shape our progressing picture of disk-induced binary orbital evolution.

\acknowledgements
The authors acknowledge funding through Harvard from ITC fellowships at the beginning of this work. DJD received funding from the European Union’s Horizon 2020 research and innovation programme under the Marie Sklodowska-Curie grant agreement No. 101029157 and through Villum Fonden grant No. 29466. We thank Zolt\'an Haiman, Andrew MacFadyen, and Jonathan Zrake for discussions that contributed to this work over a long period of time and for comments on the manuscript. We thank the anonymous referee for a constructive report, and Diego Mu\~noz, Johan Samsing and Martin Pessah for useful conversations. Numerical calculations were carried out on the Harvard Odyssey cluster.


\bibliographystyle{apj} 
\bibliography{refs}

\begin{thebibliography}{}
\expandafter\ifx\csname natexlab\endcsname\relax\def\natexlab#1{#1}\fi

\bibitem[{{Abbott} {et~al.}(2019){Abbott}, {Abbott}, {Abbott}, {Abraham},
  {Acernese}, {Ackley}, {Adams}, {Adhikari}, {Adya}, {Affeldt}, {Agathos},
  {Agatsuma}, {Aggarwal}, {Aguiar}, {Aiello}, {Ain}, {Ajith}, {Allen},
  {Allocca}, {Aloy}, {Altin}, {Amato}, {Ananyeva}, {Anderson}, {Anderson},
  {Angelova}, {Antier}, {Appert}, {Arai}, {Araya}, {Areeda}, {Ar{\`e}ne},
  {Arnaud}, {Arun}, {Ascenzi}, {Ashton}, {Aston}, {Astone}, {Aubin}, {Aufmuth},
  {AultONeal}, {Austin}, {Avendano}, {Avila-Alvarez}, {Babak}, {Bacon},
  {Badaracco}, {Bader}, {Bae}, {Baker}, {Baldaccini}, {Ballardin}, {Ballmer},
  {Banagiri}, {Barayoga}, {Barclay}, {Barish}, {Barker}, {Barkett}, {Barnum},
  {Barone}, {Barr}, {Barsotti}, {Barsuglia}, {Barta}, {Bartlett}, {Bartos},
  {Bassiri}, {Basti}, {Bawaj}, {Bayley}, {Bazzan}, {B{\'e}csy}, {Bejger},
  {Belahcene}, {Bell}, {Beniwal}, {Berger}, {Bergmann}, {Bernuzzi}, {Bero},
  {Berry}, {Bersanetti}, {Bertolini}, {Betzwieser}, {Bhandare}, {Bidler},
  {Bilenko}, {Bilgili}, {Billingsley}, {Birch}, {Birney}, {Birnholtz},
  {Biscans}, {Biscoveanu}, {Bisht}, {Bitossi}, {Bizouard}, {Blackburn},
  {Blackman}, {Blair}, {Blair}, {Blair}, {Bloemen}, {Bode}, {Boer}, {Boetzel},
  {Bogaert}, {Bondu}, {Bonilla}, {Bonnand}, {Booker}, {Boom}, {Booth}, {Bork},
  {Boschi}, {Bose}, {Bossie}, {Bossilkov}, {Bosveld}, {Bouffanais}, {Bozzi},
  {Bradaschia}, {Brady}, {Bramley}, {Branchesi}, {Brau}, {Briant}, {Briggs},
  {Brighenti}, {Brillet}, {Brinkmann}, {Brisson}, {Brockill}, {Brooks},
  {Brown}, {Brunett}, {Buikema}, {Bulik}, {Bulten}, {Buonanno}, {Buskulic},
  {Bustamante Rosell}, {Buy}, {Byer}, {Cabero}, {Cadonati}, {Cagnoli},
  {Cahillane}, {Calder{\'o}n Bustillo}, {Callister}, {Calloni}, {Camp},
  {Campbell}, {Canepa}, {Cannon}, {Cao}, {Cao}, {Capocasa}, {Carbognani},
  {Caride}, {Carney}, {Carullo}, {Casanueva Diaz}, {Casentini}, {Caudill},
  {Cavagli{\`a}}, {Cavalier}, {Cavalieri}, {Cella}, {Cerd{\'a}-Dur{\'a}n},
  {Cerretani}, {Cesarini}, {Chaibi}, {Chakravarti}, {Chamberlin}, {Chan},
  {Chao}, {Charlton}, {Chase}, {Chassande-Mottin}, {Chatterjee}, {Chaturvedi},
  {Chatziioannou}, {Cheeseboro}, {Chen}, {Chen}, {Chen}, {Cheng}, {Cheong},
  {Chia}, {Chincarini}, {Chiummo}, {Cho}, {Cho}, {Cho}, {Christensen}, {Chu},
  {Chua}, {Chung}, {Chung}, {Ciani}, {Ciobanu}, {Ciolfi}, {Cipriano}, {Cirone},
  {Clara}, {Clark}, {Clearwater}, {Cleva}, {Cocchieri}, {Coccia}, {Cohadon},
  {Cohen}, {Colgan}, {Colleoni}, {Collette}, {Collins}, {Cominsky},
  {Constancio}, {Conti}, {Cooper}, {Corban}, {Corbitt}, {Cordero-Carri{\'o}n},
  {Corley}, {Cornish}, {Corsi}, {Cortese}, {Costa}, {Cotesta}, {Coughlin},
  {Coughlin}, {Coulon}, {Countryman}, {Couvares}, {Covas}, {Cowan}, {Coward},
  {Cowart}, {Coyne}, {Coyne}, {Creighton}, {Creighton}, {Cripe}, {Croquette},
  {Crowder}, {Cullen}, {Cumming}, {Cunningham}, {Cuoco}, {Canton}, {D{\'a}lya},
  {Danilishin}, {D'Antonio}, {Danzmann}, {Dasgupta}, {Da Silva Costa},
  {Datrier}, {Dattilo}, {Dave}, {Davier}, {Davis}, {Daw}, {DeBra},
  {Deenadayalan}, {Degallaix}, {De Laurentis}, {Del{\'e}glise}, {Del Pozzo},
  {DeMarchi}, {Demos}, {Dent}, {De Pietri}, {Derby}, {De Rosa}, {De Rossi},
  {DeSalvo}, {de Varona}, {Dhurandhar}, {D{\'\i}az}, {Dietrich}, {Di Fiore},
  {Di Giovanni}, {Di Girolamo}, {Di Lieto}, {Ding}, {Di Pace}, {Di Palma}, {Di
  Renzo}, {Dmitriev}, {Doctor}, {Donovan}, {Dooley}, {Doravari}, {Dorrington},
  {Downes}, {Drago}, {Driggers}, {Du}, {Ducoin}, {Dupej}, {Dwyer}, {Easter},
  {Edo}, {Edwards}, {Effler}, {Ehrens}, {Eichholz}, {Eikenberry}, {Eisenmann},
  {Eisenstein}, {Essick}, {Estelles}, {Estevez}, {Etienne}, {Etzel}, {Evans},
  {Evans}, {Fafone}, {Fair}, {Fairhurst}, {Fan}, {Farinon}, {Farr}, {Farr},
  {Fauchon-Jones}, {Favata}, {Fays}, {Fazio}, {Fee}, {Feicht}, {Fejer}, {Feng},
  {Fernandez-Galiana}, {Ferrante}, {Ferreira}, {Ferreira}, {Ferrini},
  {Fidecaro}, {Fiori}, {Fiorucci}, {Fishbach}, {Fisher}, {Fishner},
  {Fitz-Axen}, {Flaminio}, {Fletcher}, {Flynn}, {Fong}, {Font}, {Forsyth},
  {Fournier}, {Frasca}, {Frasconi}, {Frei}, {Freise}, {Frey}, {Frey},
  {Fritschel}, {Frolov}, {Fulda}, {Fyffe}, {Gabbard}, {Gadre}, {Gaebel},
  {Gair}, {Gammaitoni}, {Ganija}, {Gaonkar}, {Garcia},
  {Garc{\'\i}a-Quir{\'o}s}, {Garufi}, {Gateley}, {Gaudio}, {Gaur}, {Gayathri},
  {Gemme}, {Genin}, {Gennai}, {George}, {George}, {Gergely}, {Germain},
  {Ghonge}, {Ghosh}, {Ghosh}, {Ghosh}, {Giacomazzo}, {Giaime}, {Giardina},
  {Giazotto}, {Gill}, {Giordano}, {Glover}, {Godwin}, {Goetz}, {Goetz},
  {Goncharov}, {Gonz{\'a}lez}, {Gonzalez Castro}, {Gopakumar}, {Gorodetsky},
  {Gossan}, {Gosselin}, {Gouaty}, {Grado}, {Graef}, {Granata}, {Grant}, {Gras},
  {Grassia}, {Gray}, {Gray}, {Greco}, {Green}, {Green}, {Gretarsson}, {Groot},
  {Grote}, {Grunewald}, {Gruning}, {Guidi}, {Gulati}, {Guo}, {Gupta}, {Gupta},
  {Gustafson}, {Gustafson}, {Haegel}, {Halim}, {Hall}, {Hall}, {Hamilton},
  {Hammond}, {Haney}, {Hanke}, {Hanks}, {Hanna}, {Hannam}, {Hannuksela},
  {Hanson}, {Hardwick}, {Haris}, {Harms}, {Harry}, {Harry}, {Haster},
  {Haughian}, {Hayes}, {Healy}, {Heidmann}, {Heintze}, {Heitmann}, {Hello},
  {Hemming}, {Hendry}, {Heng}, {Hennig}, {Heptonstall}, {Hernandez Vivanco},
  {Heurs}, {Hild}, {Hinderer}, {Hoak}, {Hochheim}, {Hofman}, {Holgado},
  {Holland}, {Holt}, {Holz}, {Hopkins}, {Horst}, {Hough}, {Howell}, {Hoy},
  {Hreibi}, {Huang}, {Huerta}, {Huet}, {Hughey}, {Hulko}, {Husa}, {Huttner},
  {Huynh-Dinh}, {Idzkowski}, {Iess}, {Ingram}, {Inta}, {Intini}, {Irwin},
  {Isa}, {Isac}, {Isi}, {Iyer}, {Izumi}, {Jacqmin}, {Jadhav}, {Jani},
  {Janthalur}, {Jaranowski}, {Jenkins}, {Jiang}, {Johnson}, {Johnson-McDaniel},
  {Jones}, {Jones}, {Jones}, {Jonker}, {Ju}, {Junker}, {Kalaghatgi},
  {Kalogera}, {Kamai}, {Kandhasamy}, {Kang}, {Kanner}, {Kapadia}, {Karki},
  {Karvinen}, {Kashyap}, {Kasprzack}, {Katsanevas}, {Katsavounidis}, {Katzman},
  {Kaufer}, {Kawabe}, {Keerthana}, {K{\'e}f{\'e}lian}, {Keitel}, {Kennedy},
  {Key}, {Khalili}, {Khan}, {Khan}, {Khan}, {Khan}, {Khazanov}, {Khursheed},
  {Kijbunchoo}, {Kim}, {Kim}, {Kim}, {Kim}, {Kim}, {Kim}, {Kimball}, {King},
  {King}, {Kinley-Hanlon}, {Kirchhoff}, {Kissel}, {Kleybolte}, {Klika},
  {Klimenko}, {Knowles}, {Koch}, {Koehlenbeck}, {Koekoek}, {Koley},
  {Kondrashov}, {Kontos}, {Koper}, {Korobko}, {Korth}, {Kowalska}, {Kozak},
  {Kringel}, {Krishnendu}, {Kr{\'o}lak}, {Kuehn}, {Kumar}, {Kumar}, {Kumar},
  {Kumar}, {Kuo}, {Kutynia}, {Kwang}, {Lackey}, {Lai}, {Lam}, {Landry}, {Lane},
  {Lang}, {Lange}, {Lantz}, {Lanza}, {Lartaux-Vollard}, {Lasky}, {Laxen},
  {Lazzarini}, {Lazzaro}, {Leaci}, {Leavey}, {Lecoeuche}, {Lee}, {Lee}, {Lee},
  {Lee}, {Lee}, {Lee}, {Lehmann}, {Lenon}, {Leroy}, {Letendre}, {Levin}, {Li},
  {Li}, {Li}, {Li}, {Lin}, {Linde}, {Linker}, {Littenberg}, {Liu}, {Liu}, {Lo},
  {Lockerbie}, {London}, {Longo}, {Lorenzini}, {Loriette}, {Lormand},
  {Losurdo}, {Lough}, {Lousto}, {Lovelace}, {Lower}, {L{\"u}ck}, {Lumaca},
  {Lundgren}, {Lynch}, {Ma}, {Macas}, {Macfoy}, {MacInnis}, {Macleod},
  {Macquet}, {Maga{\~n}a-Sandoval}, {Maga{\~n}a Zertuche}, {Magee}, {Majorana},
  {Maksimovic}, {Malik}, {Man}, {Mandic}, {Mangano}, {Mansell}, {Manske},
  {Mantovani}, {Marchesoni}, {Marion}, {M{\'a}rka}, {M{\'a}rka}, {Markakis},
  {Markosyan}, {Markowitz}, {Maros}, {Marquina}, {Marsat}, {Martelli},
  {Martin}, {Martin}, {Martynov}, {Mason}, {Massera}, {Masserot}, {Massinger},
  {Masso-Reid}, {Mastrogiovanni}, {Matas}, {Matichard}, {Matone}, {Mavalvala},
  {Mazumder}, {McCann}, {McCarthy}, {McClelland}, {McCormick}, {McCuller},
  {McGuire}, {McIver}, {McManus}, {McRae}, {McWilliams}, {Meacher}, {Meadors},
  {Mehmet}, {Mehta}, {Meidam}, {Melatos}, {Mendell}, {Mercer}, {Mereni},
  {Merilh}, {Merzougui}, {Meshkov}, {Messenger}, {Messick}, {Metzdorff},
  {Meyers}, {Miao}, {Michel}, {Middleton}, {Mikhailov}, {Milano}, {Miller},
  {Miller}, {Millhouse}, {Mills}, {Milovich-Goff}, {Minazzoli}, {Minenkov},
  {Mishkin}, {Mishra}, {Mistry}, {Mitra}, {Mitrofanov}, {Mitselmakher},
  {Mittleman}, {Mo}, {Moffa}, {Mogushi}, {Mohapatra}, {Montani}, {Moore},
  {Moraru}, {Moreno}, {Morisaki}, {Mours}, {Mow-Lowry}, {Mukherjee},
  {Mukherjee}, {Mukherjee}, {Mukund}, {Mullavey}, {Munch}, {Mu{\~n}iz},
  {Muratore}, {Murray}, {Nagar}, {Nardecchia}, {Naticchioni}, {Nayak},
  {Neilson}, {Nelemans}, {Nelson}, {Nery}, {Neunzert}, {Ng}, {Ng}, {Nguyen},
  {Nichols}, {Nielsen}, {Nissanke}, {Nitz}, {Nocera}, {North}, {Nuttall},
  {Obergaulinger}, {Oberling}, {O'Brien}, {O'Dea}, {Ogin}, {Oh}, {Oh}, {Ohme},
  {Ohta}, {Okada}, {Oliver}, {Oppermann}, {Oram}, {O'Reilly}, {Ormiston},
  {Ortega}, {O'Shaughnessy}, {Ossokine}, {Ottaway}, {Overmier}, {Owen}, {Pace},
  {Pagano}, {Page}, {Pai}, {Pai}, {Palamos}, {Palashov}, {Palomba},
  {Pal-Singh}, {Pan}, {Pang}, {Pang}, {Pankow}, {Pannarale}, {Pant},
  {Paoletti}, {Paoli}, {Papa}, {Parida}, {Parker}, {Pascucci}, {Pasqualetti},
  {Passaquieti}, {Passuello}, {Patil}, {Patricelli}, {Pearlstone}, {Pedersen},
  {Pedraza}, {Pedurand}, {Pele}, {Penn}, {Perego}, {Perez}, {Perreca},
  {Pfeiffer}, {Phelps}, {Phukon}, {Piccinni}, {Pichot}, {Piergiovanni},
  {Pillant}, {Pinard}, {Pirello}, {Pitkin}, {Poggiani}, {Pong}, {Ponrathnam},
  {Popolizio}, {Porter}, {Powell}, {Prajapati}, {Prasad}, {Prasai}, {Prasanna},
  {Pratten}, {Prestegard}, {Privitera}, {Prodi}, {Prokhorov}, {Puncken},
  {Punturo}, {Puppo}, {P{\"u}rrer}, {Qi}, {Quetschke}, {Quinonez}, {Quintero},
  {Quitzow-James}, {Raab}, {Radkins}, {Radulescu}, {Raffai}, {Raja}, {Rajan},
  {Rajbhandari}, {Rakhmanov}, {Ramirez}, {Ramos-Buades}, {Rana}, {Rao},
  {Rapagnani}, {Raymond}, {Razzano}, {Read}, {Regimbau}, {Rei}, {Reid},
  {Reitze}, {Ren}, {Ricci}, {Richardson}, {Richardson}, {Ricker},
  {Riemenschneider}, {Riles}, {Rizzo}, {Robertson}, {Robie}, {Robinet},
  {Rocchi}, {Rolland}, {Rollins}, {Roma}, {Romanelli}, {Romano}, {Romel},
  {Romie}, {Rose}, {Rosi{\'n}ska}, {Rosofsky}, {Ross}, {Rowan}, {R{\"u}diger},
  {Ruggi}, {Rutins}, {Ryan}, {Sachdev}, {Sadecki}, {Sakellariadou}, {Salafia},
  {Salconi}, {Saleem}, {Salemi}, {Samajdar}, {Sammut}, {Sanchez}, {Sanchez},
  {Sanchis-Gual}, {Sandberg}, {Sanders}, {Santiago}, {Sarin}, {Sassolas},
  {Sathyaprakash}, {Saulson}, {Sauter}, {Savage}, {Schale}, {Scheel},
  {Scheuer}, {Schmidt}, {Schnabel}, {Schofield}, {Sch{\"o}nbeck}, {Schreiber},
  {Schulte}, {Schutz}, {Schwalbe}, {Scott}, {Scott}, {Seidel}, {Sellers},
  {Sengupta}, {Sennett}, {Sentenac}, {Sequino}, {Sergeev}, {Setyawati},
  {Shaddock}, {Shaffer}, {Shahriar}, {Shaner}, {Shao}, {Sharma}, {Shawhan},
  {Shen}, {Shink}, {Shoemaker}, {Shoemaker}, {ShyamSundar}, {Siellez},
  {Sieniawska}, {Sigg}, {Silva}, {Singer}, {Singh}, {Singhal}, {Sintes},
  {Sitmukhambetov}, {Skliris}, {Slagmolen}, {Slaven-Blair}, {Smith}, {Smith},
  {Somala}, {Son}, {Sorazu}, {Sorrentino}, {Souradeep}, {Sowell}, {Spencer},
  {Srivastava}, {Srivastava}, {Staats}, {Stachie}, {Standke}, {Steer},
  {Steinke}, {Steinlechner}, {Steinlechner}, {Steinmeyer}, {Stevenson},
  {Stocks}, {Stone}, {Stops}, {Strain}, {Stratta}, {Strigin}, {Strunk},
  {Sturani}, {Stuver}, {Sudhir}, {Summerscales}, {Sun}, {Sunil}, {Suresh},
  {Sutton}, {Swinkels}, {Szczepa{\'n}czyk}, {Tacca}, {Tait}, {Talbot},
  {Talukder}, {Tanner}, {T{\'a}pai}, {Taracchini}, {Tasson}, {Taylor}, {Thies},
  {Thomas}, {Thomas}, {Thondapu}, {Thorne}, {Thrane}, {Tiwari}, {Tiwari},
  {Tiwari}, {Toland}, {Tonelli}, {Tornasi}, {Torres-Forn{\'e}}, {Torrie},
  {T{\"o}yr{\"a}}, {Travasso}, {Traylor}, {Tringali}, {Trovato}, {Trozzo},
  {Trudeau}, {Tsang}, {Tse}, {Tso}, {Tsukada}, {Tsuna}, {Tuyenbayev}, {Ueno},
  {Ugolini}, {Unnikrishnan}, {Urban}, {Usman}, {Vahlbruch}, {Vajente},
  {Valdes}, {van Bakel}, {van Beuzekom}, {van den Brand}, {Van Den Broeck},
  {Vander-Hyde}, {van Heijningen}, {van der Schaaf}, {van Veggel}, {Vardaro},
  {Varma}, {Vass}, {Vas{\'u}th}, {Vecchio}, {Vedovato}, {Veitch}, {Veitch},
  {Venkateswara}, {Venugopalan}, {Verkindt}, {Vetrano}, {Vicer{\'e}}, {Viets},
  {Vine}, {Vinet}, {Vitale}, {Vo}, {Vocca}, {Vorvick}, {Vyatchanin}, {Wade},
  {Wade}, {Wade}, {Walet}, {Walker}, {Wallace}, {Walsh}, {Wang}, {Wang},
  {Wang}, {Wang}, {Wang}, {Ward}, {Warden}, {Warner}, {Was}, {Watchi},
  {Weaver}, {Wei}, {Weinert}, {Weinstein}, {Weiss}, {Wellmann}, {Wen},
  {Wessel}, {We{\ss}els}, {Westhouse}, {Wette}, {Whelan}, {White}, {Whiting},
  {Whittle}, {Wilken}, {Williams}, {Williamson}, {Willis}, {Willke}, {Wimmer},
  {Winkler}, {Wipf}, {Wittel}, {Woan}, {Woehler}, {Wofford}, {Worden},
  {Wright}, {Wu}, {Wysocki}, {Xiao}, {Yamamoto}, {Yancey}, {Yang}, {Yap},
  {Yazback}, {Yeeles}, {Yu}, {Yu}, {Yuen}, {Yvert}, {Zadro{\.Z}ny}, {Zanolin},
  {Zappa}, {Zelenova}, {Zendri}, {Zevin}, {Zhang}, {Zhang}, {Zhang}, {Zhao},
  {Zhou}, {Zhou}, {Zhu}, {Zimmerman}, {Zlochower}, {Zucker}, {Zweizig}, {LIGO
  Scientific Collaboration}, \& {Virgo Collaboration}}]{LIGO_GWTC1:2019}
{Abbott}, B.~P., {Abbott}, R., {Abbott}, T.~D., {et~al.} 2019, Physical Review
  X, 9, 031040

\bibitem[{{Alves} {et~al.}(2019){Alves}, {Caselli}, {Girart}, {Segura-Cox},
  {Franco}, {Schmiedeke}, \& {Zhao}}]{Alves+2019}
{Alves}, F.~O., {Caselli}, P., {Girart}, J.~M., {et~al.} 2019, Science, 366, 90

\bibitem[{{Aly} {et~al.}(2015){Aly}, {Dehnen}, {Nixon}, \&
  {King}}]{AlyNixon+2015}
{Aly}, H., {Dehnen}, W., {Nixon}, C., \& {King}, A. 2015, \mnras, 449, 65

\bibitem[{{Amaro-Seoane} \& et~al.(2017)}]{LISA:2017}
{Amaro-Seoane}, P., \& et~al. 2017, ArXiv e-prints, arXiv:1702.00786

\bibitem[{{Armitage} \& {Natarajan}(2002)}]{ArmNat:2002}
{Armitage}, P.~J., \& {Natarajan}, P. 2002, \apjl, 567, L9

\bibitem[{{Arzoumanian} {et~al.}(2020){Arzoumanian}, {Baker}, {Blumer},
  {B{\'e}csy}, {Brazier}, {Brook}, {Burke-Spolaor}, {Chatterjee}, {Chen},
  {Cordes}, {Cornish}, {Crawford}, {Cromartie}, {Decesar}, {Demorest}, {Dolch},
  {Ellis}, {Ferrara}, {Fiore}, {Fonseca}, {Garver-Daniels}, {Gentile}, {Good},
  {Hazboun}, {Holgado}, {Islo}, {Jennings}, {Jones}, {Kaiser}, {Kaplan},
  {Kelley}, {Key}, {Laal}, {Lam}, {Lazio}, {Lorimer}, {Luo}, {Lynch},
  {Madison}, {McLaughlin}, {Mingarelli}, {Ng}, {Nice}, {Pennucci}, {Pol},
  {Ransom}, {Ray}, {Shapiro-Albert}, {Siemens}, {Simon}, {Spiewak}, {Stairs},
  {Stinebring}, {Stovall}, {Sun}, {Swiggum}, {Taylor}, {Turner}, {Vallisneri},
  {Vigeland}, {Witt}, \& {Nanograv Collaboration}}]{NANOGrav12p5_2020}
{Arzoumanian}, Z., {Baker}, P.~T., {Blumer}, H., {et~al.} 2020, \apjl, 905, L34

\bibitem[{{Bate}(2000)}]{Bate+2000}
{Bate}, M.~R. 2000, \mnras, 314, 33

\bibitem[{{Begelman} {et~al.}(1980){Begelman}, {Blandford}, \&
  {Rees}}]{Begel:Blan:Rees:1980}
{Begelman}, M.~C., {Blandford}, R.~D., \& {Rees}, M.~J. 1980, \nat, 287, 307

\bibitem[{{Dempsey} {et~al.}(2020){Dempsey}, {Mu{\~n}oz}, \&
  {Lithwick}}]{Dempsey+2020}
{Dempsey}, A.~M., {Mu{\~n}oz}, D., \& {Lithwick}, Y. 2020, \apjl, 892, L29

\bibitem[{{Dittmann} \& {Ryan}(2021)}]{DittRyan:2021}
{Dittmann}, A., \& {Ryan}, G. 2021, arXiv e-prints, arXiv:2102.05684

\bibitem[{{D'Orazio} {et~al.}(2015){D'Orazio}, {Haiman}, {Duffell}, {Farris},
  \& {MacFadyen}}]{PG1302MNRAS:2015a}
{D'Orazio}, D.~J., {Haiman}, Z., {Duffell}, P., {Farris}, B.~D., \&
  {MacFadyen}, A.~I. 2015, \mnras, 452, 2540

\bibitem[{D'Orazio {et~al.}(2016)D'Orazio, Haiman, Duffell, MacFadyen, \&
  Farris}]{D'Orazio:CBDTrans:2016}
D'Orazio, D.~J., Haiman, Z., Duffell, P., MacFadyen, A., \& Farris, B. 2016,
  MNRAS, 459, 2379

\bibitem[{{D'Orazio} {et~al.}(2013){D'Orazio}, {Haiman}, \&
  {MacFadyen}}]{DHM:2013}
{D'Orazio}, D.~J., {Haiman}, Z., \& {MacFadyen}, A. 2013, \mnras, 436, 2997

\bibitem[{{Duffell}(2016)}]{DuffellDISCO:2016}
{Duffell}, P.~C. 2016, \apjs, 226, 2

\bibitem[{{Duffell} {et~al.}(2020){Duffell}, {D'Orazio}, {Derdzinski},
  {Haiman}, {MacFadyen}, {Rosen}, \& {Zrake}}]{Duffell+2020}
{Duffell}, P.~C., {D'Orazio}, D., {Derdzinski}, A., {et~al.} 2020, \apj, 901,
  25

\bibitem[{{Dunhill} {et~al.}(2015){Dunhill}, {Cuadra}, \&
  {Dougados}}]{Dunhill+2015}
{Dunhill}, A.~C., {Cuadra}, J., \& {Dougados}, C. 2015, \mnras, 448, 3545

\bibitem[{{El-Badry} {et~al.}(2019){El-Badry}, {Rix}, {Tian}, {Duch{\^e}ne}, \&
  {Moe}}]{El-BadryTwins+2019}
{El-Badry}, K., {Rix}, H.-W., {Tian}, H., {Duch{\^e}ne}, G., \& {Moe}, M. 2019,
  \mnras, 489, 5822

\bibitem[{{Farris} {et~al.}(2014){Farris}, {Duffell}, {MacFadyen}, \&
  {Haiman}}]{Farris+2014}
{Farris}, B.~D., {Duffell}, P., {MacFadyen}, A.~I., \& {Haiman}, Z. 2014, \apj,
  783, 134

\bibitem[{{Gerosa} {et~al.}(2015){Gerosa}, {Veronesi}, {Lodato}, \&
  {Rosotti}}]{Gerosa_spinalign+2015}
{Gerosa}, D., {Veronesi}, B., {Lodato}, G., \& {Rosotti}, G. 2015, \mnras, 451,
  3941

\bibitem[{{Goodchild} \& {Ogilvie}(2006)}]{GoodchildOgilvie:2006}
{Goodchild}, S., \& {Ogilvie}, G. 2006, \mnras, 368, 1123

\bibitem[{{Gould} \& {Rix}(2000)}]{GouldRix:2000}
{Gould}, A., \& {Rix}, H.-W. 2000, \apjl, 532, L29

\bibitem[{{Haiman} {et~al.}(2009){Haiman}, {Kocsis}, \& {Menou}}]{HKM09}
{Haiman}, Z., {Kocsis}, B., \& {Menou}, K. 2009, \apj, 700, 1952

\bibitem[{{Heath} \& {Nixon}(2020)}]{HeathNixon:2020}
{Heath}, R.~M., \& {Nixon}, C.~J. 2020, \aap, 641, A64

\bibitem[{{Jensen} \& {Akeson}(2014)}]{JensenAkeson:2014}
{Jensen}, E. L.~N., \& {Akeson}, R. 2014, \nat, 511, 567

\bibitem[{{Kelley} {et~al.}(2019){Kelley}, {Haiman}, {Sesana}, \&
  {Hernquist}}]{Kelley+2019}
{Kelley}, L.~Z., {Haiman}, Z., {Sesana}, A., \& {Hernquist}, L. 2019, \mnras,
  485, 1579

\bibitem[{{Li} {et~al.}(2021){Li}, {Dempsey}, {Li}, {Li}, \&
  {Li}}]{LiDempsey+2021}
{Li}, Y.-P., {Dempsey}, A.~M., {Li}, S., {Li}, H., \& {Li}, J. 2021, \apj, 911,
  124

\bibitem[{{Lubow}(1991)}]{Lubow:1991}
{Lubow}, S.~H. 1991, \apj, 381, 259

\bibitem[{MacFadyen \& Milosavljevi{\'c}(2008)}]{MM08}
MacFadyen, A.~I., \& Milosavljevi{\'c}, M. 2008, \apj, 672, 83

\bibitem[{{Martin}(2019)}]{DMartin_CBplanets:2019}
{Martin}, D.~V. 2019, \mnras, 488, 3482

\bibitem[{{Mathieu} {et~al.}(1997){Mathieu}, {Stassun}, {Basri}, {Jensen},
  {Johns-Krull}, {Valenti}, \& {Hartmann}}]{Mathieu_DQTau+1997}
{Mathieu}, R.~D., {Stassun}, K., {Basri}, G., {et~al.} 1997, \aj, 113, 1841

\bibitem[{{Miranda} {et~al.}(2017){Miranda}, {Mu{\~n}oz}, \&
  {Lai}}]{MirandaLai+2017}
{Miranda}, R., {Mu{\~n}oz}, D.~J., \& {Lai}, D. 2017, \mnras, 466, 1170

\bibitem[{{Moody} {et~al.}(2019){Moody}, {Shi}, \& {Stone}}]{MoodyStone:2019}
{Moody}, M. S.~L., {Shi}, J.-M., \& {Stone}, J.~M. 2019, \apj, 875, 66

\bibitem[{{M{\"o}sta} {et~al.}(2019){M{\"o}sta}, {Taam}, \&
  {Duffell}}]{Moesta+2019}
{M{\"o}sta}, P., {Taam}, R.~E., \& {Duffell}, P.~C. 2019, \apjl, 875, L21

\bibitem[{{Mu{\~n}oz} \& {Lai}(2016)}]{MunozLai+2016}
{Mu{\~n}oz}, D.~J., \& {Lai}, D. 2016, \apj, 827, 43

\bibitem[{{Mu{\~n}oz} {et~al.}(2020){Mu{\~n}oz}, {Lai}, {Kratter}, \&
  {Miranda}}]{Munoz_FinDsk+2020}
{Mu{\~n}oz}, D.~J., {Lai}, D., {Kratter}, K., \& {Miranda}, R. 2020, \apj, 889,
  114

\bibitem[{{Mu{\~n}oz} \& {Lithwick}(2020)}]{MunozLithwick:2020}
{Mu{\~n}oz}, D.~J., \& {Lithwick}, Y. 2020, \apj, 905, 106

\bibitem[{{Mu{\~n}oz} {et~al.}(2019){Mu{\~n}oz}, {Miranda}, \&
  {Lai}}]{MunozLai+2019}
{Mu{\~n}oz}, D.~J., {Miranda}, R., \& {Lai}, D. 2019, \apj, 871, 84

\bibitem[{{M{\"u}ller} {et~al.}(2012){M{\"u}ller}, {Kley}, \&
  {Meru}}]{Mueller2DhydroGS+2012}
{M{\"u}ller}, T.~W.~A., {Kley}, W., \& {Meru}, F. 2012, \aap, 541, A123

\bibitem[{{Nixon} {et~al.}(2013){Nixon}, {King}, \& {Price}}]{Nixon+2013}
{Nixon}, C., {King}, A., \& {Price}, D. 2013, \mnras, 434, 1946

\bibitem[{{Price-Whelan} {et~al.}(2020){Price-Whelan}, {Hogg}, {Rix}, {Beaton},
  {Lewis}, {Nidever}, {Almeida}, {Badenes}, {Barba}, {Beers}, {Carlberg}, {De
  Lee}, {Fern{\'a}ndez-Trincado}, {Frinchaboy}, {Garc{\'\i}a-Hern{\'a}ndez},
  {Green}, {Hasselquist}, {Longa-Pe{\~n}a}, {Majewski}, {Nitschelm}, {Sobeck},
  {Stassun}, {Stringfellow}, \& {Troup}}]{AMPW_APOGEEII+2020}
{Price-Whelan}, A.~M., {Hogg}, D.~W., {Rix}, H.-W., {et~al.} 2020, \apj, 895, 2

\bibitem[{{Roedig} {et~al.}(2011){Roedig}, {Dotti}, {Sesana}, {Cuadra}, \&
  {Colpi}}]{Roedig+2011}
{Roedig}, C., {Dotti}, M., {Sesana}, A., {Cuadra}, J., \& {Colpi}, M. 2011,
  \mnras, 415, 3033

\bibitem[{{Roedig} {et~al.}(2014){Roedig}, {Krolik}, \& {Miller}}]{Roedig+2014}
{Roedig}, C., {Krolik}, J.~H., \& {Miller}, M.~C. 2014, \apj, 785, 115

\bibitem[{{Shi} {et~al.}(2012){Shi}, {Krolik}, {Lubow}, \&
  {Hawley}}]{ShiKrolik:2012}
{Shi}, J.-M., {Krolik}, J.~H., {Lubow}, S.~H., \& {Hawley}, J.~F. 2012, \apj,
  749, 118

\bibitem[{{Smallwood} {et~al.}(2021){Smallwood}, {Martin}, \&
  {Lubow}}]{Smallwood+2021}
{Smallwood}, J.~L., {Martin}, R.~G., \& {Lubow}, S.~H. 2021, \apjl, 907, L14

\bibitem[{{Tagawa} {et~al.}(2020){Tagawa}, {Haiman}, \& {Kocsis}}]{Tagawa+2020}
{Tagawa}, H., {Haiman}, Z., \& {Kocsis}, B. 2020, \apj, 898, 25

\bibitem[{{Tang} {et~al.}(2017){Tang}, {MacFadyen}, \& {Haiman}}]{Tang+2017}
{Tang}, Y., {MacFadyen}, A., \& {Haiman}, Z. 2017, \mnras, 469, 4258

\bibitem[{{Thun} {et~al.}(2017){Thun}, {Kley}, \& {Picogna}}]{Thun+2017}
{Thun}, D., {Kley}, W., \& {Picogna}, G. 2017, \aap, 604, A102

\bibitem[{{Tiede} {et~al.}(2020){Tiede}, {Zrake}, {MacFadyen}, \&
  {Haiman}}]{Tiede+2020}
{Tiede}, C., {Zrake}, J., {MacFadyen}, A., \& {Haiman}, Z. 2020, \apj, 900, 43

\bibitem[{{Tofflemire} {et~al.}(2017){Tofflemire}, {Mathieu}, {Herczeg},
  {Akeson}, \& {Ciardi}}]{Tofflemire_TWA3A+2017}
{Tofflemire}, B.~M., {Mathieu}, R.~D., {Herczeg}, G.~J., {Akeson}, R.~L., \&
  {Ciardi}, D.~R. 2017, \apjl, 842, L12

\bibitem[{{Zrake} {et~al.}(2021){Zrake}, {Tiede}, {MacFadyen}, \&
  {Haiman}}]{Zrake+2020}
{Zrake}, J., {Tiede}, C., {MacFadyen}, A., \& {Haiman}, Z. 2021, \apjl, 909,
  L13

\end{thebibliography}
\end{document}